\documentclass[aps,prx,twocolumn,superscriptaddress,]{revtex4-1}
\usepackage{amsmath}
\usepackage{amsfonts}
\usepackage{graphicx}
\usepackage{color}
\usepackage{dsfont}
\usepackage{verbatim}
\usepackage{mathtools}
\usepackage[normalem]{ulem}
\usepackage{comment}
\usepackage{stackrel}
\usepackage{bm}
\usepackage[pdftex]{hyperref}
\hypersetup{colorlinks = true, urlcolor = black, linkcolor = black, citecolor = black}

\definecolor{myblue}{rgb}{.93, .93, 1}

\newcommand{\bsub}{\begin{subequations}}
	\newcommand{\esub}{\end{subequations}}

\newcommand{\vex}[1]{\bm{\mathrm{#1}}}

\begin{document}
	
	\title{Acoustic-phonon-mediated superconductivity in moir\'eless graphene multilayers}
	
	\author{Yang-Zhi~Chou}\email{yzchou@umd.edu}
	\affiliation{Condensed Matter Theory Center and Joint Quantum Institute, Department of Physics, University of Maryland, College Park, Maryland 20742, USA}
	
	\author{Fengcheng~Wu}
	\affiliation{School of Physics and Technology, Wuhan University, Wuhan 430072, China}
	\affiliation{Wuhan Institute of Quantum Technology, Wuhan 430206, China}
	
	\author{Jay D. Sau}
	\affiliation{Condensed Matter Theory Center and Joint Quantum Institute, Department of Physics, University of Maryland, College Park, Maryland 20742, USA}
	
	\author{Sankar Das~Sarma}
	\affiliation{Condensed Matter Theory Center and Joint Quantum Institute, Department of Physics, University of Maryland, College Park, Maryland 20742, USA}	
	\date{\today}
	
\begin{abstract}
We investigate the competition between acoustic-phonon-mediated superconductivity and the long-range Coulomb interaction in moir\'eless graphene multilayers, specifically, Bernal bilayer graphene, rhombohedral trilayer graphene, and ABCA-stacked tetralayer graphene. In these graphene multilayers, the acoustic phonons can realize, through electron-phonon coupling, both spin-singlet and spin-triplet pairings, and the intra-sublattice pairings ($s$-wave spin-singlet and $f$-wave spin-triplet) are the dominant channels. Our theory naturally explains the distinct recent experimental findings in Bernal bilayer graphene and rhombohedral trilayer graphene, and we further predict the existence of superconductivity in ABCA tetralayer graphene arising from electron-phonon interactions. In particular, we demonstrate that the acoustic-phonon-mediated superconductivity prevails over a wide range of doping in rhombohedral trilayer graphene and ABCA tetralayer graphene while superconductivity exists only in a narrow range of doping near the Van Hove singularity in Bernal bilayer graphene. Key features of our theory are the inclusion of realistic band structures with the appropriate Van Hove singularities and Coulomb repulsion effects (the so-called ``$\mu^*$ effect'') opposing the phonon-induced superconducting pairing. We also discuss how intervalley scatterings can suppress the spin-triplet spin-polarized superconductivity. Our work provides detailed prediction based on electron-acoustic-phonon-interaction-induced graphene superconductivity, which should be investigated in future experiments. 
\end{abstract}
	
	\maketitle

\section{Introduction}

Superconductivity is a prominent and extensively-studied quantum many-body phenomenon because of its fundamental importance, widespread occurrence in nature, and technological applications. One of the most active contemporary research directions in condensed matter physics is the superconductivity in magic-angle moir\'e graphene systems including magic-angle twisted bilayer graphene \cite{Cao2018_tbg1,Cao2018_tbg2,Yankowitz2019,Lu2019}, magic-angle twisted trilayer graphene \cite{Hao2021electric,Park2021tunable,Cao2021,Liu2021coulomb}, and magic-angle twisted graphene multilayers ($n>3$) \cite{Zhang2021ascendance,Park2021MAMG,Burg2022emergence}. The single-particle bands in such systems are tuned to be nearly flat \cite{Bistritzer,Li2019,Khalaf2019} such that many-body effects can become significant. It is also worth mentioning that robust reproducible superconductivity has not been systematically established in other moir\'e systems, making magic-angle twisted graphene systems distinctive. In addition, the extensively studied regular monolayer graphene is not known to be superconducting (because the electron-phonon coupling \cite{Hwang2008,Efetov2010} is not significant enough to produce an observable $T_c$ for a doped monolayer graphene), adding considerable excitement to the unexpected discovery of superconductivity in moir\'e magic angle twisted graphene layers. 

Twisting or a	 moir\'e flatband or magic angle, however, is not an essential condition for superconductivity in graphene-based materials as rhombohedral trilayer graphene (RTG) \cite{Zhou2021,Zhou2021_SC_RTG} and Bernal bilayer graphene (BBG) \cite{Zhou2021_BBG} also demonstrate robust superconducting behavior in recent experimental studies. There are two distinct superconducting phases in RTG, termed SC1 and SC2. The superconductivity in SC1 is suppressed by an in-plane magnetic field within the Pauli limit, which is thought to be more consistent with a spin-singlet pairing (although complicated spin-triplet pairing/singlet-triplet mixing could in some situations also manifest similar physics); SC2 is likely to be a (spin-polarized) spin-triplet pairing since the superconductivity persists under a large in-plane magnetic field violating the Pauli limit. By contrast, superconductivity in BBG is rather mysterious -- a sufficiently large in-plane magnetic field is required to induce a (spin-polarized) spin-triplet superconducting state in BBG. Since similar spin-singlet/spin-triplet superconductivity has been observed experimentally in magic-angle moir\'e graphene systems \cite{Cao2018_tbg1,Cao2018_tbg2,Yankowitz2019,Lu2019,Hao2021electric,Park2021tunable,Cao2021,Liu2021coulomb,Zhang2021ascendance,Park2021MAMG,Burg2022emergence}, a reasonable question is whether there exists a universal pairing mechanism for superconductivity in all graphene-based materials, with and without moir\'e structure. %Although, in principle, it is possible that the observed superconductivity in various graphene multilayers, twisted and untwisted, arises from different underlying mechanisms, simple Occam's razor consideration implies this to be unlikely as it would be akin to different superconducting metals (e.g. Al, Pb, Ag, Cu) somehow having different underlying mechanisms. 
In this context, acoustic phonons are the most natural candidate for pairing since Cooper pairings in most superconductors in nature are caused by acoustic phonons, and in the untwisted systems with no moir\'e flat bands, the most obvious arguments in favor of strong correlation induced superconductivity become questionable. In the current work, we develop a detailed theory for acoustic phonon-induced superconductivity in `moir\'eless' graphene multilayers, where no twist is involved between the layers.

Before discussing a potential universal mechanism, it is important to emphasize that superconductivity in graphene-based materials is distinct from other systems (e.g., conventional metals) because of the valley and sublattice degrees of freedom. For example, electron-acoustic-phonon coupling in graphene has an enlarged $SU(2)\times SU(2)$ symmetry due to the approximate valley symmetry \cite{Wu2019_phonon}. As a result, the acoustic-phonon-mediated intervalley pairings have a singlet-triplet degeneracy, and the intrasublattice pairings are typically favored \cite{Wu2019_phonon,Chou2021correlation}. Therefore, it is natural to ask whether acoustic-phonon-mediated pairings can account for superconductivity in graphene-based materials \cite{Wu2019_phonon,Wu2020_TDBG}. Previously, we showed that the acoustic-phonon-mediated superconductivity can explain qualitatively and semi-quantitatively the distinct superconducting phenomenology reported in RTG \cite{Chou2021_RTG_SC} and BBG \cite{Chou2021_BBG}. Since the band structures of RTG and BBG are simpler and better established compared with twisted moir\'e graphene systems, it is easier to make direct (semi-)quantitative comparisons between theory and experiment here. Besides the acoustic-phonon mechanism, which we consider, a number of alternative theoretical ideas focusing on inter-electron interactions have also been proposed for RTG \cite{Chatterjee2021,Ghazaryan2021,Dong2021,Cea2021,Szabo2021,You2021,Qin2022,Dai2022} and BBG \cite{Szabo2021BBG}. We mention here that the case for robust acoustic phonon mediated superconductivity in twisted graphene systems has already been made in the literature, based on the enhancement of the effective electron-phonon coupling in moir\'e systems by virtue of the suppression of the graphene Fermi velocity \cite{Wu2019_phonon,Wu2020_TDBG}, but the current work, by contrast, is specifically on moir\'eless graphene multilayers.

In this work, we investigate in considerable details the acoustic-phonon-mediated superconductivity in moir\'eless graphene multilayers including BBG, RTG, and (the experimentally not-yet-studied, and thus, we are making a prediction) ABCA stacked tetralayer graphene \cite{Kerelsky2021moireless}. We incorporate the $\vex{k}\cdot\vex{p}$ band structure and the Coulomb repulsion in our phonon-induced theory of superconductivity. For RTG and ABCA tetralayer graphene, we find that robust observable superconductivity ($T_c>20$mK) can be realized for a wide range of doping, even for doping away from the Van Hove singularity (VHS) while observable and rather fragile superconductivity is obtained only near VHS in BBG. Thus, we predict the existence of a more generic doping-independent (and also more robust) superconductivity in RTG and ABCA than in BBG. Our work, while being in agreement with the existing experimental observations, also provides a number of falsifiable predictions based on electron-acoustic-phonon coupling incorporating Coulomb repulsion, and we believe, based on our finding a reasonable agreement between our theory and experiment, that acoustic phonons are the main mediators of superconductivity for graphene-based materials in general. A unique feature of graphene is the fact that acoustic phonons can lead to both singlet and triplet superconductivity because of the enlarged $SU(2)\times SU(2)$ symmetry enabled by the valley degrees of freedom. 
%The fact that graphene superconductivity has been observed both in moir\'e and in moir\'eless multilayers also argues in favor of acoustic phonon effects rather than correlation effects since the flatband enhancement of correlations is absent in the non-moir\'e untwisted multilayers.

The rest of the paper is organized as follows: In Sec.~\ref{Sec:Model}, we introduce the $\vex{k}\cdot\vex{p}$ band model, the electron-acoustic-phonon coupling, and the Coulomb interaction. We discuss how to incorporate Coulomb repulsion in the theory of acoustic-phonon-mediated graphene superconductivity and present a simplified approach in Sec.~\ref{Sec:SC}. In Sec.~\ref{Sec:Results}, the main numerical results are presented and discussed in the context of experimental results. We conclude with a brief discussion in Sec.~\ref{Sec:Discussion}. Appendices A-F complement the main text by providing various technical details used in our work.

\section{Microscopic Model}\label{Sec:Model}

Superconductivity is crucially dependent on density of states (DOS) and microscopic interactions (e.g., electron-phonon, electron-electron). In this section, we discuss the single-particle band structures, and interactions used in this work. We focus on untwisted moir\'eless pristine BBG, RTG, and ABCA-stacked tetralayer graphene.

\begin{figure}[t!]
	\includegraphics[width=0.45\textwidth]{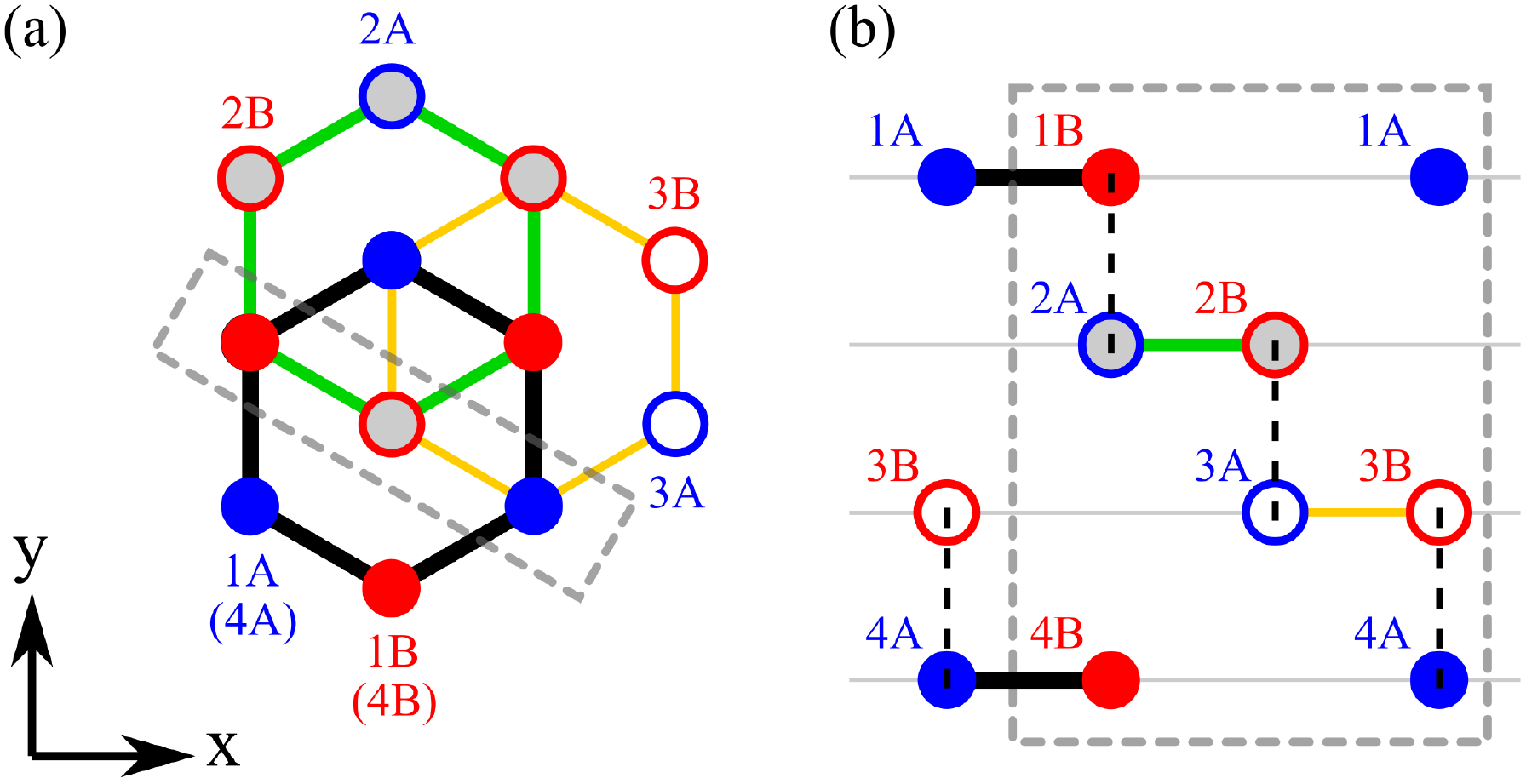}
	\caption{Lattice structure of ABCA-stacked tetralayer graphene. (a) Top view. For each layer, we illustrate a hexagon to specify the relative position in the $xy$ plane. 1A, 2A, 3A, and 4A (1B, 2B, 3B, and 4B) denote the sublattice A (B) in the layer 1, 2, 3, and 4 respectively. Note that the lattice points in the first layer and the fourth layer are at the same $xy$ positions. (b) The cross-section view. At K and $-$K points, the intra-layer hybridizations can be ignored, and the nearest neighbor inter-layer couplings generate dimerization in 1B-2A, 2B-3A, 3B-4A bonds (black dashed bonds). 1A and 3B sites are the low-energy sites in this simplified picture. The BBG (RTG) structure can be derived from ABCA tetralayer graphene with only the first 2 (3) layers.
	}
	\label{Fig:ABCA}
\end{figure}

\subsection{Single-particle band structure}

We are interested, following the experimental systems, in the low-doping graphene multilayers in the presence of a displacement field, which induces a tunable band gap at the charge neutrality point. The single-particle bands near the $K$ and $-K$ valleys can be described by $\vex{k}\cdot\vex{p}$ band models. Generally, the single-particle Hamiltonian is described by
\begin{align}\label{Eq:H_lowE}
	\hat{H}_{n,0}=\sum_{\tau}\sum_{\vex{k}}\hat{\Psi}_{n,\tau}^{\dagger}(\vex{k})\hat{h}_{n,\tau}(\vex{k})\hat{1}_{s}\hat{\Psi}_{n,\tau}(\vex{k}),
\end{align}
where $\hat{h}_{n,\tau=\pm}(\vex{k})$ is a $2n\times 2n$ low-energy Hamiltonian near the $\pm K$ valley, $n\ge 2$ is the number of layers, $\hat{1}_{s}$ is the identity matrix in the spin space, and $\hat{\Psi}_{n,\tau}(\vex{k})$ is a $4n$-component column vector with a valley quantum number $\tau$, made of the fermionic annihilation operator $\psi_{\tau\sigma l s}$ with sublattice $\sigma$, spin $s$, and layer $l$. In this work, we consider BBG ($n=2$), RTG ($n=3$), and ABCA tetralayer graphene ($n=4$).

The low-energy bands of the graphene multilayer systems here have large probability on the A sites of the top layer (1A) and B sites of the bottom layer ($n$B). This property arises from the interlayer nearest-neighbor tunnelings which tend to form dimerized bonds as illustrated in Fig.~\ref{Fig:ABCA}, where ABCA tetralayer graphene is illustrated. One can obtain BBG (RTG) by considering just two (three) layers in Fig.~\ref{Fig:ABCA}. To gain some intuitive understanding, we construct an effective $2\times 2$ matrix given by \cite{Zhang2010}
\begin{align}\label{Eq:h_two_band}
	\hat{h}_{n,+}'(\vex{k})\approx\left[\begin{array}{cc}
		\Delta_1	& C_n\left(\Pi_k^{\dagger}\right)^n\\
		C_n\left(\Pi_k\right)^n	& -\Delta_1
	\end{array}
	\right],
\end{align}
where $2\Delta_1$ corresponds to the energy difference between two low-energy sites induced by the displacement field, $C_n=v_0^n/\gamma_1^{n-1}$, $v_0$ is the graphene velocity, and $\gamma_1$ corresponds to the interlayer dimerization energy. The effective energy bands are described by $\mathcal{E}_{\vex{k}}'=\pm \sqrt{C_n^2|\vex{k}|^{2n}+\Delta_1^2}$, resulting in a divergent DOS $\rho(\mathcal{E})\propto |\mathcal{E}\pm \Delta_1|^{-1+1/n}$ near the band edge ($\pm \Delta_1$). Based on this heuristic estimate, we expect that the DOS gets larger for a higher layer number ($n$). More careful analysis should include additional hopping terms and crystal fields, which might substantially alter the results, such as inducing VHS away from band edges. Regardless of the detail, the low-energy bands are approximately layer and sublattice polarized. As a result, the superconducting states with intralayer intersublattice pairings should be generically suppressed in the low-energy bands because one of the sublattices in each layer has higher energy.

To obtain the low-energy band structure, we formally diagonalize the Hamiltonian in Eq.~(\ref{Eq:H_lowE}) as follows:
\begin{align}\label{Eq:H_0_diagonalized}
	\hat{H}_0=\sum_{\tau=\pm}\sum_{b=1}^{2n}\sum_{s=\uparrow,\downarrow}\mathcal{E}_{\tau,b}(\vex{k})c^{\dagger}_{\tau b s}(\vex{k})c_{\tau b s}(\vex{k}),
\end{align}
where $\mathcal{E}_{\tau,b}(\vex{k})$ encodes the energy-momentum dispersion of the $b$th band and valley $\tau K$, and $c_{\tau b s}(\vex{k})$ is an electron annihilation operator of valley $\tau K$, the $b$th band, spin $s$, and momentum $\vex{k}$. The microscopic-basis operator $\psi_{\tau\sigma l s}$ and the band-basis operator $c_{\tau b s}$ obey $\psi_{\tau\sigma l s}(\vex{k})=\sum_b\Phi_{\tau b, \sigma l}(\vex{k})c_{\tau b s}(\vex{k})$, where $\Phi_{\tau b, \sigma l}(\vex{k})$ is the wavefunction of valley $\tau$K and band $b$. In addition, the (spinless) time-reversal symmetry imposes further constraints: $\mathcal{E}_{+,b}(\vex{k})=\mathcal{E}_{-,b}(-\vex{k})$ and $\Phi_{+ b, \sigma l}(\vex{k})=\Phi_{- b, \sigma l}^*(-\vex{k})$. 

We use the $\vex{k}\cdot\vex{p}$ bands described in Appendix~\ref{App:Bands} and compute DOS numerically for BBG, RTG, and ABCA tetralayer graphene as shown in Fig.~\ref{Fig:DOS}. (See Appendix~\ref{App:Numerics} for a discussion on the numerical calculations.) Note that the DOS in BBG is much smaller than the DOS in RTG and ABCA tetralayer graphene. The differences in DOS imply that the screened Coulomb interaction might show different behavior since screening depends crucially on the DOS, as will be discussed in detail later.

\begin{figure}[t!]
	\includegraphics[width=0.4\textwidth]{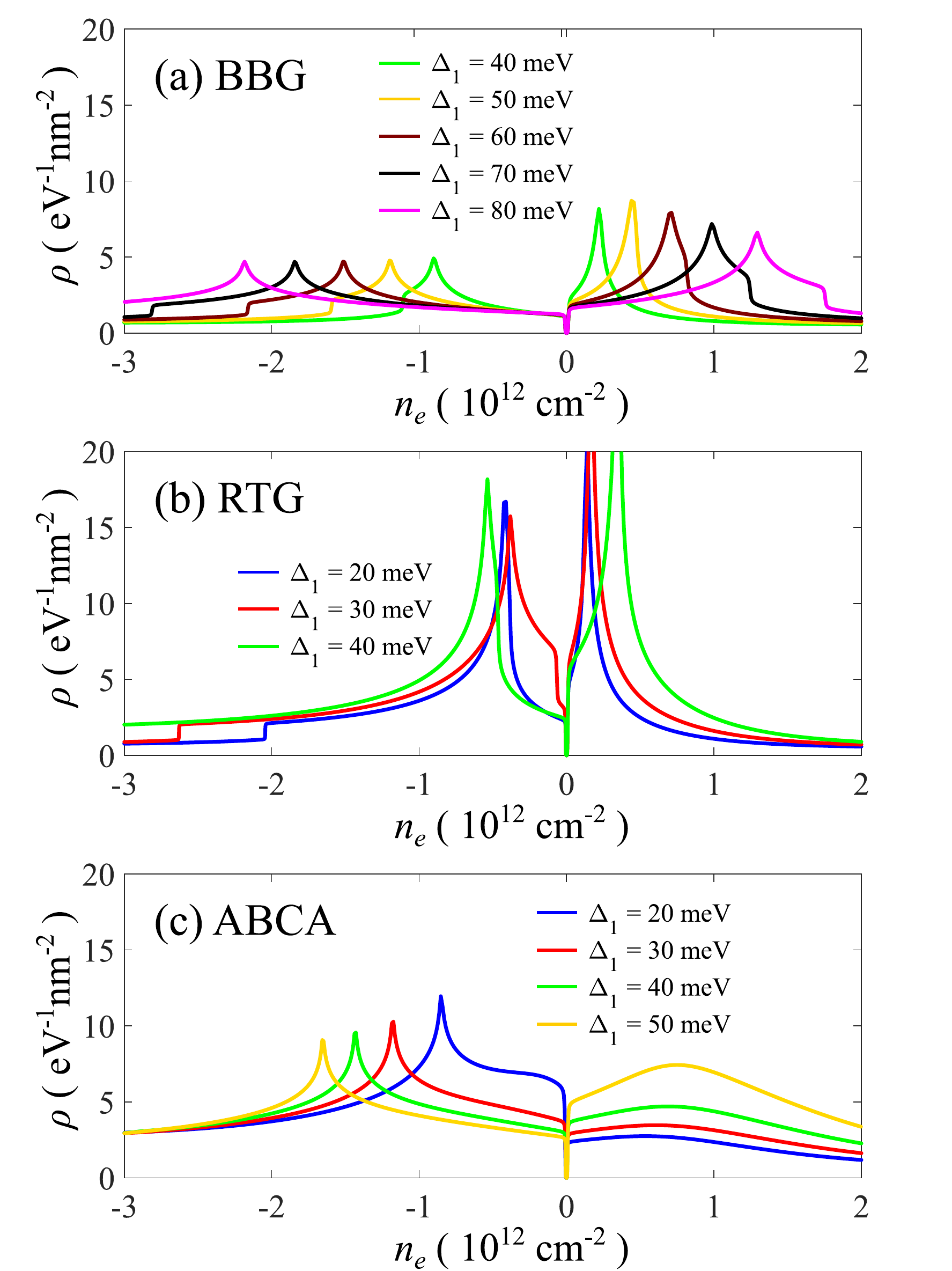}
	\caption{Density of states based on $\vex{k}\cdot\vex{p}$ models. (a) BBG (b) RTG (c) ABCA. The numerical results are obtained by computing a $10^4\times 10^4$ momentum grid with a momentum spacing $\Delta k\approx 2\times 10^{-5}a_0^{-1}$. For a given $\mathcal{E}_F$, we determine $n_e$ based on the DOS profiles illustrated here. $\Delta_1$ is a band parameter (defined in Appendix~\ref{App:Bands}) which can be tuned by a displacement field. 
	}
	\label{Fig:DOS}
\end{figure}

\subsection{Electron-phonon coupling}

In graphene multilayers, the electron-optical-phonon couplings \cite{Wu2018} are generically suppressed because of the sublattice polarization in the systems \cite{Choi2019}. Thus, we focus on the in-plane longitudinal acoustic phonon, which is described by
\begin{align}
	\hat{H}_{ph}=\sum_l\sum_{\vex{q}}\omega_{\vex{q}}a^{\dagger}_{l,\vex{q}}a_{l,\vex{q}},
\end{align}
where $a_{\vex{q}}$ is the phonon annihilation operator with momentum $\vex{q}$, $\omega_{\vex{q}}=v_{ph}|\vex{q}|$ is the acoustic phonon dispersion, and $v_{ph}$ is the sound velocity. For simplicity, we consider that the acoustic phonon modes are layer decoupled, i.e., the same as that in monolayer graphene \cite{Hwang2008}. However, our qualitative results do not rely on this assumption.

The electron-acoustic phonon coupling \cite{Coleman2015introduction} is given by, within the well-known deformation potential coupling approximation, 
\begin{align}\label{Eq:H_ep}
	\hat{H}_{ep}=\frac{D}{\sqrt{\mathcal{A}}}\sum_{\vex{q},l}\sqrt{\frac{\hbar}{2\rho_m\omega_{\vex{q}}}}\left(-i\vex{q}\cdot\hat{e}_{\vex{q}}\right)\left(a_{l,\vex{q}}+a^{\dagger}_{l,-\vex{q}}\right)\hat{n}_{l}(-\vex{q}),
\end{align}
where $\hat{e}_{\vex{q}}^*=\hat{e}_{-\vex{q}}$ is the polarization vector, $\rho_m$ is the mass density of monolayer graphene, $D$ is the deformation potential, $\mathcal{A}$ is the area of the 2D system, $\omega_{\vex{q}}=v_{\text{ph}}|\vex{q}|$ is the acoustic phonon dispersion, and $\hat{n}_l(-\vex{q})=\sum_{\vex{k}}\sum_{\tau,\sigma,s}\psi^{\dagger}_{\tau,l,\sigma,s}(\vex{k})\psi_{\tau,l,\sigma,s}(\vex{k}-\vex{q})$.

\subsection{Coulomb interaction}

In addition to electron-acoustic-phonon couplings, the electrons are directly interacting via Coulomb repulsion, which is an important factor in determining the existence of superconductivity, because no superconductivity would be possible if the repulsive Coulomb coupling overwhelms the attractive interaction induced by acoustic phonons in the associated pairing channel.
We focus on the long-range component of the instantaneous Coulomb interaction, described by
\begin{align}\label{Eq:H_C}
	\hat{H}_C=&\frac{1}{2\mathcal{A}}\sum_{\vex{q}}V_C(\vex{q})\sum_{l}\hat{n}_{l}(\vex{q})\sum_{l'}\hat{n}_{l'}(-\vex{q})
\end{align}
where $V_C(\vex{q})$ encodes the Coulomb potential and $\hat{n}_l(\vex{q})=\sum_{\vex{k}}\sum_{\tau,\sigma,s}\psi^{\dagger}_{\tau,l,\sigma,s}(\vex{k})\psi_{\tau,l,\sigma,s}(\vex{k}+\vex{q})$. The long-range Coulomb potential given by Eq.~(\ref{Eq:H_C}) has an $SU(2)\times SU(2)$ symmetry, which results in a singlet-triplet degeneracy.
However, the short-range contributions of the Coulomb potential might break the $SU(2)\times SU(2)$ symmetry down to a $SU(2)$ symmetry \cite{Chatterjee2021}.

In the experiments, the graphene multilayer system is sandwiched between two metallic plates which screen the Coulomb interaction. After solving the electrostatic problem using the image charge approximations, we obtain
\begin{align}
	V_{C}(\vex{q})=&\frac{2\pi e^2}{\epsilon |\vex{q}|}\tanh\left(\left|\vex{q}\right|d\right),
\end{align}
where $\epsilon$ is the dimensionless average background lattice dielectric constant and $d$ is the distance between the 2D system and the metallic plates. 

In addition to the gate screening, the large DOS in graphene multilayers (Fig.~\ref{Fig:DOS}) result in significant intraband screenings. To incorporate the intraband screening by the carriers themselves, we adopt the extensively-used Thomas-Fermi approximation defined by 
\begin{align}
	\label{Eq:V_C_TF}V_{\text{TF}}(\vex{q},\mathcal{E}_F)=\frac{1}{\left[V_C(\vex{q})\right]^{-1}+\rho(\mathcal{E}_F)}=\frac{V_C(\vex{q})}{1+V_C(\vex{q})\rho(\mathcal{E}_F)}
\end{align}
where $\rho(\mathcal{E}_F)$ is the total DOS at Fermi energy. The Thomas-Fermi approximation is the static limit of the random phase approximation, which is exact under the well-controlled limits of high density and/or many fermion flavors. When $V_{\text{C}}(\vex{q})\rho(\mathcal{E}_F)\gg 1$, $V_{\text{TF}}(\vex{q};\mathcal{E}_F)\approx 1/\rho(\mathcal{E}_F)$, which is independent of $\epsilon$ and $d$ (simply because, in this large DOS limit, the screening by the carriers themselves dominate). In graphene multilayers discussed here, the intraband process is the dominating mechanism for the screening of Coulomb repulsion. We will discuss the interplay between phonon-mediated pairings and screened Coulomb repulsion next.

\section{Phonon-Mediated Superconductivity incorporating Coulomb repulsion}\label{Sec:SC}

To achieve a more quantitative understanding, it is customary to apply the Eliashberg theory \cite{Marsiglio2020eliashberg,Chubukov2020eliashberg} with the full frequency dependence of the problem, which is typically solved by intensive numerical methods. This is because the retarded effective attraction can overcome the instantaneous Coulomb repulsion even though the bare interaction is repulsive at all frequencies \cite{MorelAnderson1962,Coleman2015introduction,Marsiglio2020eliashberg,Chubukov2020eliashberg}. 

In the rest of this section, we present a simplified treatment without carrying out intensive numerics, incorporating both the acoustic-phonon attraction and Coulomb repulsion, to solve for superconductivity in moir\'eless graphene multilayers. (The full numerical solution is presented in Appendix~\ref{App:Elia_numerics}.) We first discuss the effective BCS interaction and examine the retardation effect by comparing the phonon velocity with the estimated Fermi velocity, Then, we review the Eliashberg theory and present a simplified mean-field approach. We support our results by solving the Eliashberg theory numerically in Appendix \ref{App:Elia_numerics} and Fig.~\ref{Fig:Eliash}, where the qualitative agreement with our simplified almost-analytical mean-field solution is shown.

\subsection{BCS superconductivity and retardation}

Electrons near the Fermi surface can attract each other via a phonon-mediated interaction. Such an attractive interaction can overcome the repulsive Coulomb repulsion and create Cooper pairs. This is the central idea of the BCS theory. To derive the phonon-mediated attraction, we start with the electron-phonon couplings [given by Eq.~(\ref{Eq:H_ep})] and integrate out the phonon fields in the imaginary-time path integral. The effective interaction is described by an action,
\begin{align}\label{Eq:S_ph}
	\mathcal{S}_{\text{ph}}=-\frac{1}{2\beta\mathcal{A}}\sum_{\nu_n,\vex{q}}V_g(\nu_n,\vex{q})\sum_{l}\hat{n}_{l}(\nu_n,\vex{q})\hat{n}_{l}(-\nu_n,-\vex{q}),
\end{align}
where $\nu_n$ is the Matsubara frequency, $V_g(\nu_n,\vex{q})=g\frac{\omega_{\vex{q}}^2}{\omega_{\vex{q}}^2+\nu_n^2}$ is the phonon-mediated dynamical potential, $\omega_{\vex{q}}=v_s|\vex{q}|$, $v_s$ is the sound velocity, and $g=D^2/(\rho_mv_{s}^2)$ is the strength of phonon-mediated attraction. The overall minus sign indicates the effective attraction mediated by acoustic phonons, and the effective attraction has an $SU(2)\times SU(2)$ symmetry, resulting in a singlet-triplet degeneracy in the pairing. To estimate $g_0$, we use $D=30$ eV, $\rho_m=7.6\times 10^{-8}$ g/cm$^2$ \cite{Efetov2010,Hwang2008}, and $v_s=2\times 10^6$ cm/s. We obtain $g\approx474$ meV$\cdot$nm$^2$ \cite{Wu2020_TDBG,Chou2021_RTG_SC}. Here, $D=30$eV is based on the experimentally extracted value \cite{Efetov2010}, and it might be off by a factor of $2$ \cite{Wu2019_phonon,Hwang2008}. In this work, we use $g=g_0\equiv474$ meV$\cdot$nm$^2$ unless noted otherwise. Our qualitative results are independent of the choice of $D$ and $g$.

\subsubsection{Single-band approximation and pairing symmetry}

To simplify the calculations, we adopt the single-band approximation to where the Fermi energy $\mathcal{E}_F$ lies. This is a valid approximation keeping only the band because low-energy bands of the graphene multilayers are separated by a gap $\sim 2|\Delta_1|$ due to the applied displacement field, and the high energy bands are also away by at least $\sim 100$ meV. The BCS channel of the phonon-mediated interaction [Eq.~(\ref{Eq:S_ph})] is given by
\begin{align}
\label{Eq:S_ph_b}\mathcal{S}_{\text{ph}}=\frac{-1}{\beta\mathcal{A}}\sum_{k,k'}V_g^{(b)}(k,k')\bar{c}_{+bs,k}\bar{c}_{-bs',-k}c_{-bs',-k'}c_{+bs,k'},
\end{align}
where
\begin{align}
\label{Eq:V_ph_bar}V_g^{(b)}(k,k')=&g_{\vex{k},\vex{k}'}^{(b)}\frac{\omega_{\vex{k}-\vex{k}'}^2}{\omega_{\vex{k}-\vex{k}'}^2+\left(\omega_n-\omega_n'\right)^2},\\
\label{Eq:g_kk'}g_{\vex{k},\vex{k}'}^{(b)}=&g\sum_{\sigma,l}\left|\Phi_{+b;l\sigma}(\vex{k})\right|^2\left|\Phi_{+b;l\sigma}(\vex{k}')\right|^2,
\end{align}
$b$ is the index of the projected band, $c_{\tau b s,k}$, $\bar{c}_{\tau b s,k}$ are the Grassmann variables representing the fermionic fields, $k=(\omega_n,\vex{k})$ denotes the frequency-momentum index, and $V_g^{(b)}(k,k')$ is the phonon-mediated BCS attractive potential after the single-band projection.

Before we proceed, it is worthwhile discussing the pairing symmetry in the low-energy bands of graphene multilayers. We consider only the intervalley Cooper pairs here because $\mathcal{E}_{\tau,b}(\vex{k})\neq \mathcal{E}_{\tau,b}(-\vex{k})$ generically suppresses the intravalley superconductivity \cite{Einenkel2011,Sun2021}. Following the classification scheme based on valley and sublattice degrees of freedom \cite{Wu2019_phonon,Wu2020_TDBG,Chou2021correlation,Chou2021_RTG_SC}, the intervalley pairing symmetry (i.e., $s$-, $p$-, $d$-, $f$- wave) can be determined from $\mathcal{C}_{3z}$ (threefold rotation about hexagon center) and spin SU(2) symmetry. $s$-wave spin-singlet and $f$-wave spin-triplet pairings are intrasublattice; $p$-wave spin-triplet and $d$-wave spin-singlet are intersublattice. For graphene multilayers, we find that the intralayer intersublattice pairings are strongly suppressed in the low-energy bands since one of the sublattices in each layer is at high energy. Thus, we focus only on the intralayer intrasublattice pairings, i.e., $s$-wave spin-singlet and $f$-wave spin-triplet pairings. In fact, $s$-wave spin-singlet and $f$-wave spin-triplet pairings are degenerate due to the $SU(2)\times SU(2)$ symmetry in the acoustic-phonon-mediated attraction.

\subsubsection{Mean-field approximation}

In the standard BCS approximation, the frequency dependence is suppressed completely. As such, $V_g^{(b)}(k,k')$ is reduced to $g_{\vex{k},\vex{k}'}^{(b)}$. With the mean-field approximation, we derive the linearized gap equation as follows:
\begin{align}
	\label{Eq:LGE_1}\Delta_{s's}(\vex{k})=&\frac{1}{\mathcal{A}}\sum_{\vex{k}'}g_{\vex{k},\vex{k}'}^{(b)}\frac{\tanh\left[\frac{\mathcal{E}_{+b}(\vex{k}')-\mathcal{E}_F}{2k_BT}\right]}{2\mathcal{E}_{+b}(\vex{k}')-2\mathcal{E}_F}\Delta_{s's}(\vex{k}'),
\end{align}
where $k_B$ is the Boltzmann constant, $\mathcal{E}_F$ is the Fermi energy, and the superconducting order parameter is defined by
\begin{align}
	\label{Eq:Delta}\Delta_{s's}(\vex{k}')=&\frac{1}{\mathcal{A}}\sum_b\sum_{\vex{k}'}g_{\vex{k},\vex{k}'}^{(b)}\left\langle c_{-bs'}(-\vex{k}')c_{+bs}(\vex{k}')\right\rangle.
\end{align}
The transition temperature $T_c$ is determined by the highest $T$ such that Eq.~(\ref{Eq:LGE_1}) is satisfied. The obtained $T_c$ here is for both the $s$-wave spin-singlet and $f$-wave spin-triplet pairings because of the singlet-triplet degeneracy in the acoustic-phonon-mediated pairing.

\subsubsection{Validity of BCS approximation}

\begin{figure}[t!]
	\includegraphics[width=0.4\textwidth]{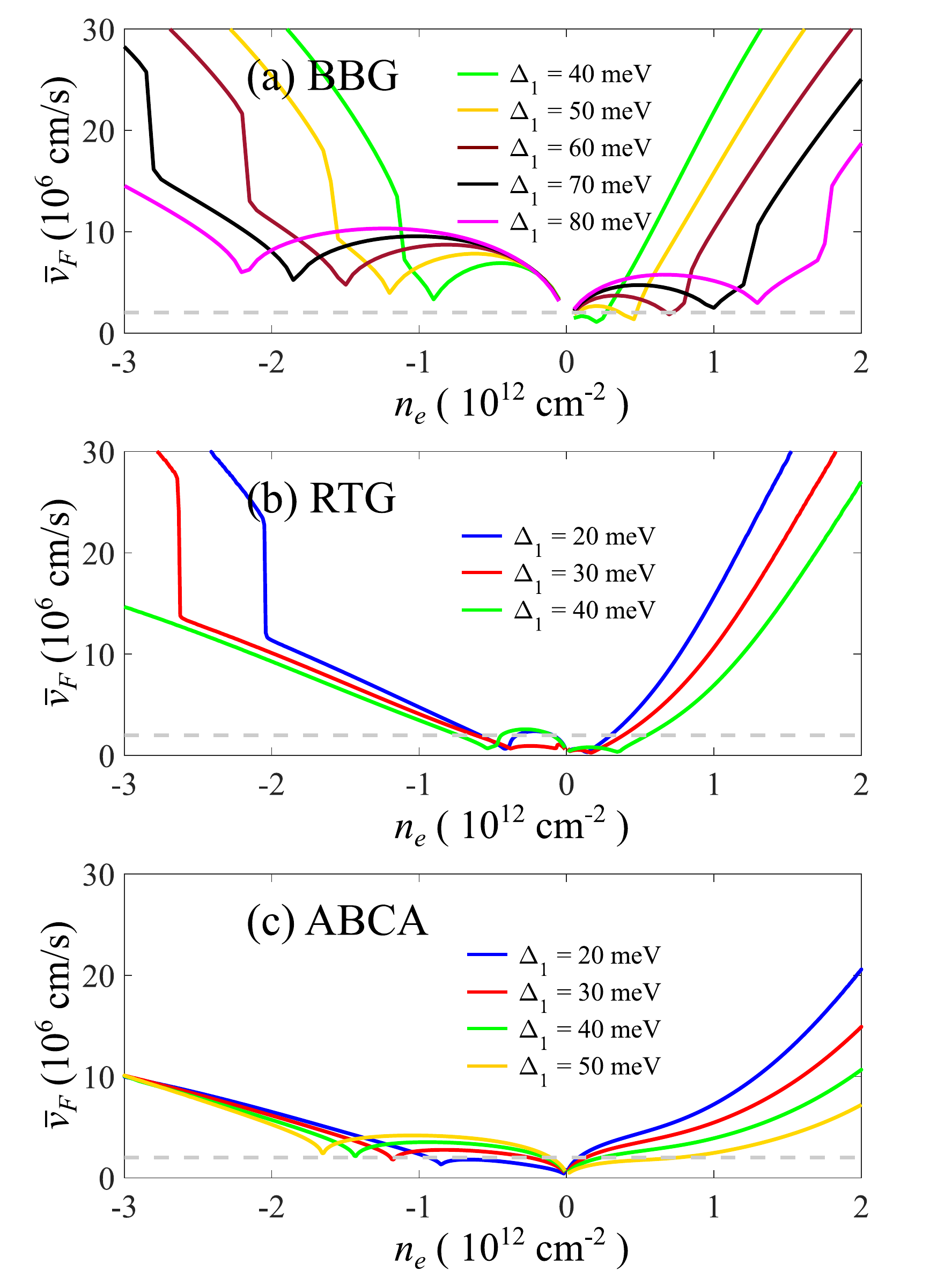}
	\caption{Estimate of averaged Fermi velocity $\bar{v}_F$ based on $\vex{k}\cdot\vex{p}$ bands. We use $\bar{v}_F=2\sqrt{|n_e|}/(\hbar\sqrt{\pi}\rho)$. (a) BBG (b) RTG (c) ABCA stacked tetralayer graphene. $\Delta_1$ is a band parameter (defined in Appendix~\ref{App:Bands}) which can be tuned by a displacement field. 
	}
	\label{Fig:vF}
\end{figure}

The validity of BCS theory relies on the retardation effect, indicating that phonon velocity is smaller than electron velocity. In such a case, the Migdal theorem applies, and vertex corrections can be ignored. However, the graphene multilayer systems contain VHS in the low energy bands, which can result in a small Fermi velocity, and our theory incorporating the Migdal theorem would break down for $v_s$ exceeding the Femi velocity. To check this, we estimate the average Fermi velocity, $\bar{v}_F=2\sqrt{|n_e|}/(\hbar\sqrt{\pi}\rho)$, where $n_e$ is the carrier density and $\rho$ is the total DOS (incorporating spin and valley, assuming unpolarized states). In Fig.~\ref{Fig:vF}, we find that $\bar{v}_F$ is larger than the sound velocity $v_s$ (gray dashed line) at generic dopings, suggesting the validity of the Migdal theorem and BCS approximation holding generally in BBG, RTG, and ABCA tetralayer graphene. For doping densities with $\bar{v}_F<v_s$ (e.g., near VHS), the non-adiabatic vertex corrections \cite{Cappelluti1996,Phan_2020} become important, and $T_c$ is generically suppressed by these vertex corrections except for situations that are deep in the anti-adiabatic limit \cite{Phan_2020}. In particular, the vertex correction for doping densities close to the VHS can increase $T_c$ \cite{Cappelluti1996}. We neglect all vertex corrections in the current work. For a fixed $|\Delta_1|$, one can see that the $\bar{v}_F$ away from VHS gets smaller for a larger $n$ (number of layers). This property is consistent with the effective two-band model description in Eq.~(\ref{Eq:h_two_band}), where dispersion is approximately proportional to $|\vex{k}|^{2n}$ near the band edge.

\subsubsection{Superconductivity without Coulomb repulsion}

We numerically solve Eq.~(\ref{Eq:LGE_1}) and plot $T_c$ versus $n_e$ in Fig.~\ref{Fig:BCS} for BBG, RTG, and ABCA tetralayer graphene. The numerical parameters are provided in Appendix~\ref{App:Numerics}. The results show that observable $T_c$ is produced for a wide range of doping for all systems, suggesting that acoustic phonons can induce superconductivity in these systems. We emphasize that $T_c$ is determined by a wide window of energy states near $\mathcal{E}_F$ but not just the states precisely at $\mathcal{E}_F$ \cite{Lothman2017,Chou2021_RTG_SC}. Thus, the Fermi energy precisely being at the VHS is not crucial for the emergent superconductivity. Technically, this is due to the kernel $\frac{\tanh\left[x/T\right]}{2x}$ in Eq.~(\ref{Eq:LGE_1}) having a finite width, which has a power-law falling off in $x$ for $x\gg T$. This is distinct from the Stoner-type instability where the kernel is reduced to a Dirac-delta function at $T=0$. Typically, the $T_c$ values predicted in Fig.~\ref{Fig:BCS}, without any Coulomb repulsion effects, overestimate the actual $T_c$ because Coulomb effects suppress $T_c$. To provide quantitative predictions, Coulomb repulsion has to be incorporated. Next, we turn to a framework incorporating both the phonon-mediated attraction and Coulomb repulsion.

\begin{figure}[t!]
	\includegraphics[width=0.45\textwidth]{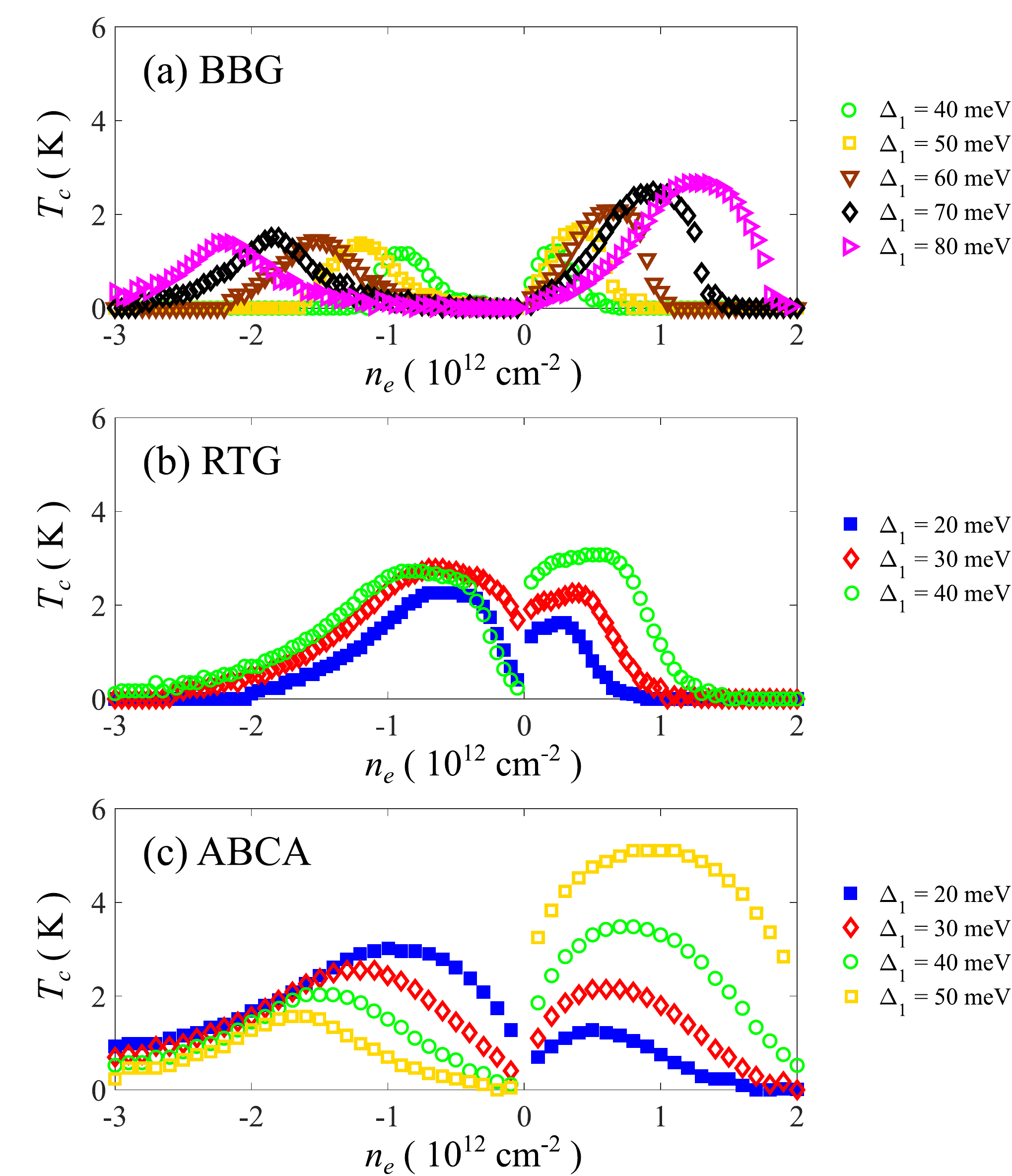}
	\caption{Numerical $T_c$ based on pure electron-acoustic-phonon pairing. (a) BBG (b) RTG (c) ABCA. We sole Eq.~(\ref{Eq:LGE_1}) with 5000 energy levels from a fine momentum grid with a spacing $\Delta k \approx0.002 a_0^{-1}$. (a) is the same as Ref.~\cite{Chou2021_BBG}. (b) is slightly different from Ref.~\cite{Chou2021_RTG_SC} (in the low doping) due to the finer momentum mesh and the way of determining $n_e$. $\Delta_1$ is a band parameter (defined in Appendix~\ref{App:Bands}) which can be tuned by a displacement field. 
	}
	\label{Fig:BCS}
\end{figure}

\subsection{Eliashberg theory and renormalization of Coulomb interaction}

To investigate the interplay between phonon-mediated attraction and direct Coulomb repulsion, the frequency dependence, which is ignored in BCS theory, should be taken into account. We review the celebrated Eliashberg theory \cite{Marsiglio2020eliashberg,Chubukov2020eliashberg} within the single-band approximation (projection onto the $b$th band) in the following. There are two sets of equations in Eliashberg theory, a self-consistent equation for determining the Eliashberg self energy, and another self-consistent equation for determining the order parameter. See Appendix~\ref{App:Elia_Th} for a derivation based on the path integral. The main results are summarized in the following. 

\subsubsection{Eliashberg equations}

We focus on $T\approx T_c$ where the order parameter is infinitesimal.
In such a situation, the Eliashberg self energy is determined by 
\begin{align}\label{Eq:Eliashberg_Xi}
i\Xi_{+s}(k')=&\frac{1}{\beta \mathcal{A}}\sum_{k}\frac{- W(k',k)}{-i\omega_n+\mathcal{E}_{+,b}(\vex{k})-\mathcal{E}_F+i\Xi_{+s}(k)},
\end{align}
where $i\Xi_{+s}(k)$ is the Eliashberg self energy of valley $+K$, spin $s$, $W(k,k')=V^{(b)}_g(k,k')-V^{(b)}_{\text{TF}}(k,k')$, and $V^{(b)}_{\text{TF}}(k,k')$ denotes Eq.~(\ref{Eq:V_C_TF}) after projecting onto the $b$th band. The Eliashberg self energy can be written as $i\Xi_{+s}(k)=\left(-Z_k+1\right)i\omega_n+\chi_k$, where $Z_k$ is the wavefunction renormalization and $\chi_k$ encodes the dispersion renormalization and quasiparticle life time. Using the Eliashberg self energy, the linearized gap equation is expressed as
\begin{align}\label{Eq:LGE_Eliash}
\Delta_{ss'}(k')=\frac{1}{\beta\mathcal{A}}\sum_{k}\frac{W(k',k)\Delta_{ss'}(k)}{\left(Z_k\omega_n\right)^2+\left[\mathcal{E}_{+,b}(\vex{k})-\mathcal{E}_F+\chi_k\right]^2},
\end{align}
where we have ignored the infinitesimal $|\Delta_{ss'}(k)|$ term in the denominator. 
To simply the calculations, we set $Z_k=1$ and ignore $\chi_k$. Equation (\ref{Eq:LGE_Eliash}) becomes a frequency-dependent BCS gap equation given by
\begin{align}\label{Eq:LGE_freq}
\Delta_{ss'}(k')=\frac{1}{\beta\mathcal{A}}\sum_{k}\frac{W(k',k)\Delta_{ss'}(k)}{\omega_n^2+\left[\mathcal{E}_{+,b}(\vex{k})-\mathcal{E}_F\right]^2}.
\end{align}
This approximation is valid in the weak electron-phonon coupling limit, which certainly applied to the multilayer graphene systems under consideration in the current work. Our qualitative results do not rely on this assumption. Note that Eq.~(\ref{Eq:LGE_freq}) is reduced to the frequency-independent BCS gap equation Eq.~(\ref{Eq:LGE_1}) after suppressing the frequency dependence in $\Delta$ and $\tilde{W}$.

\begin{figure}[t!]
	\includegraphics[width=0.45\textwidth]{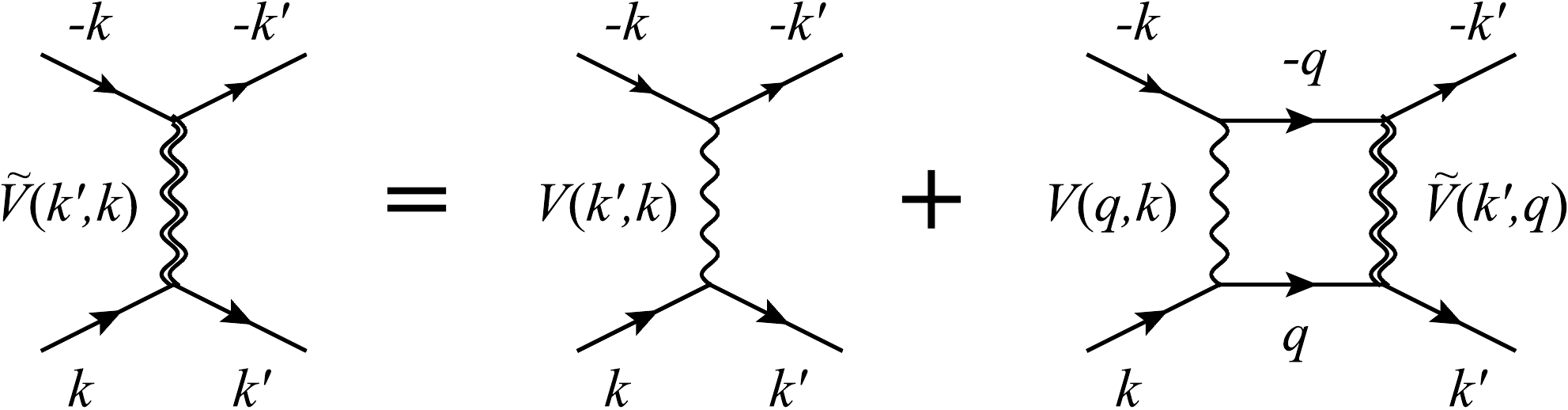}
	\caption{Diagrammatic representation of self-consistent ladder equation for renormalization from high energy states. The single wiggly lines denote the bare interaction $V	$; the double wiggly lines denote the renormalized interaction $\tilde{V}$; the solid lines with arrows denote the electron propagators. Note that $k$, $k'$, and $p$ are in valley $K$ while $-k$, $-k'$, and $-p$ are in valley $-K$.
	}
	\label{Fig:Ladder}
\end{figure}

\subsubsection{Frequency-dependent gap equation}

Solving the integral equation defined by Eq.~(\ref{Eq:LGE_Eliash}) or (\ref{Eq:LGE_freq}) is a highly challenging computational task since a large momentum mesh and a large frequency mesh are required. Our goal here is to map the frequency-dependent Eliashberg theory into an effective frequency-independent BCS theory incorporating the so-called $\mu^*$ effect of Coulomb repulsion. To derive the $\mu^*$ effect, we first assume that the phonon-mediated attraction is only nonzero around the Fermi level in the low frequency regime ($|\omega_n|,|\omega_n'|<\omega_c$) while the Coulomb repulsion is essentially frequency independent. Now, we rewrite the frequency-dependent BCS gap equation [given by Eq.~(\ref{Eq:LGE_freq})] as follows:
\begin{align}\label{Eq:LGE_two_regime}
\Delta(k)=\frac{1}{\beta\mathcal{A}}\sum_{k'}\left[\chi_C(\vex{k};\vex{k}')+\chi_{\text{ph}}(k;k')\right]\Delta(k'),
\end{align}
where 
\begin{align}
\Delta(k)=&\Theta(\omega_c-|\omega_n|)\Delta_{<}(k)+\Theta(|\omega_n|-\omega_c)\Delta_{>}(k),\\
\chi_C(k;k')=&-V^{(b)}_{\text{TF}}(k,k')\frac{1}{\omega_n'^2+\left[\mathcal{E}_{+,b}(\vex{k})-\mathcal{E}_F\right]^2},\\
\chi_{\text{ph}}(k;k')=&V^{(b)}_g(k,k')\frac{\Theta(\omega_c-|\omega_n|)\Theta(\omega_c-|\omega_n'|)}{\omega_n'^2+\left[\mathcal{E}_{+,b}(\vex{k})-\mathcal{E}_F\right]^2}.
\end{align}
If we assume that the high frequency ($|\omega_n|>\omega_c$) gap function does not depend on $\omega_n$, i.e., $\Delta_>(k)=\Delta_{\infty}(\vex{k})$, equation (\ref{Eq:LGE_two_regime}) reduces to two coupled equations as follows:
\begin{align}
	\nonumber\Delta_<(k)=&\frac{1}{\beta\mathcal{A}}\sum_{k',|\omega_n'|<\omega_c}\left[\chi_C(k,k')+\chi_{\text{ph}}(k,k')\right]\Delta_{<}(k')\\
\label{Eq:Delta_<}&+\frac{1}{\beta\mathcal{A}}\sum_{k',|\omega_n'|>\omega_c}\chi_C(k,k')\Delta_{\infty}(\vex{k}'),\text{ for }|\omega_n|<\omega_c,\\
	\nonumber\Delta_{\infty}(\vex{k})=&\frac{1}{\beta\mathcal{A}}\sum_{k',|\omega_n'|<\omega_c}\chi_C(k,k')\Delta_{<}(k')\\
\label{Eq:Delta_inf}&+\frac{1}{\beta\mathcal{A}}\sum_{k',|\omega_n'|>\omega_c}\chi_C(k,k')\Delta_{\infty}(\vex{k}'),\text{ for }|\omega_n|>\omega_c.
\end{align}
Formally, one can eliminate $\Delta_{\infty}(\vex{k})$ and derive an effective gap equation \cite{Marsiglio2020eliashberg} as follows:
\begin{align}\label{Eq:LGE_eff}
	\Delta_<(k)=&\frac{1}{\beta\mathcal{A}}\sum_{k',|\omega_n'|<\omega_c}\left[\tilde{\chi}_C(k,k')+\chi_{\text{ph}}(k,k')\right]\Delta_{<}(k'),
\end{align}
where $\tilde{\chi}_C(k,k')$ encodes the Coulomb repulsion after integrating out the high frequency degrees of freedom. Note that $\chi_{\text{ph}}(k,k')$ is unchanged during this process as $\chi_{\text{ph}}(k,k')=0$ in the high frequency regime. 

\subsubsection{$\mu^*$ effect}

The renormalization from the high energy states can also reduce the Coulomb repulsion in the BCS channel, which we treat by solving the ladder self-consistent equation \cite{Coleman2015introduction} shown in Fig.~\ref{Fig:Ladder}. The self-consistent ladder Dyson equation corresponds to an algebraic equation as follows:
\begin{align}\label{Eq:DysonE}
\tilde{V}(k',k)=V(k',k)-\frac{1}{\beta \mathcal{A}}\sum_{\substack{\nu_n,\vex{q},\\
		\omega_c<|\nu_n|<\Lambda,\\
	|\tilde{\mathcal{E}}_{\vex{q}}|<\Lambda}}\frac{\tilde{V}(k',q)V(q,k)}{\nu_n^2+\tilde{\mathcal{E}}_{\vex{q}}},
\end{align}
where $\tilde{\mathcal{E}}_{\vex{q}}=\mathcal{E}_+(\vex{q})-\mathcal{E}_F$, $\tilde{V}$ is the renormalized interaction, and $V$ is the bare interaction. This is equivalent to deriving $\tilde{\chi}_C$ in Eq.~(\ref{Eq:LGE_eff}). If we ignore the momentum dependence of the screened Coulomb interaction and use $U_0(\mathcal{E}_F)\equiv V_{\text{TF}}(k_F;\mathcal{E}_F)$, the renormalized interaction is given by
\begin{align}\label{Eq:U_R}
	U_R(\mathcal{E}_F)=\frac{U_0(\mathcal{E}_F)}{1+U_0(\mathcal{E}_F)\Gamma(\mathcal{E}_F;\omega_c;\Lambda)},
\end{align}
where 
$\Gamma(\mathcal{E}_F;\omega_c;\Lambda)$ encodes the renormalization from the energies satisfying $\omega_c<|\mathcal{E}_{\tau b}(\vex{k})-\mathcal{E}_F|<\Lambda$, $\omega_c=2v_sk_F$, and $\Lambda$ is the energy cutoff. We discuss how to numerically evaluate $\Gamma$ for arbitrary band structures in Appendix~\ref{App:Gamma}. If we assume a constant DOS ($\rho_0$), the well-established $\mu^*$ formula \cite{Coleman2015introduction} is reproduced,
\begin{align}
	\mu^*=\frac{\mu}{1+\mu\ln(\Lambda/\omega_c)},
\end{align}
where $\mu^*=U_R(\mathcal{E}_F) \rho_0$ and $\mu=U_0(\mathcal{E}_F) \rho_0$ are the dimensionless renormalized and bare interaction, respectively.

\subsection{BCS theory with effective attraction}\label{Sec:g_star}

\begin{figure}[t!]
	\includegraphics[width=0.4\textwidth]{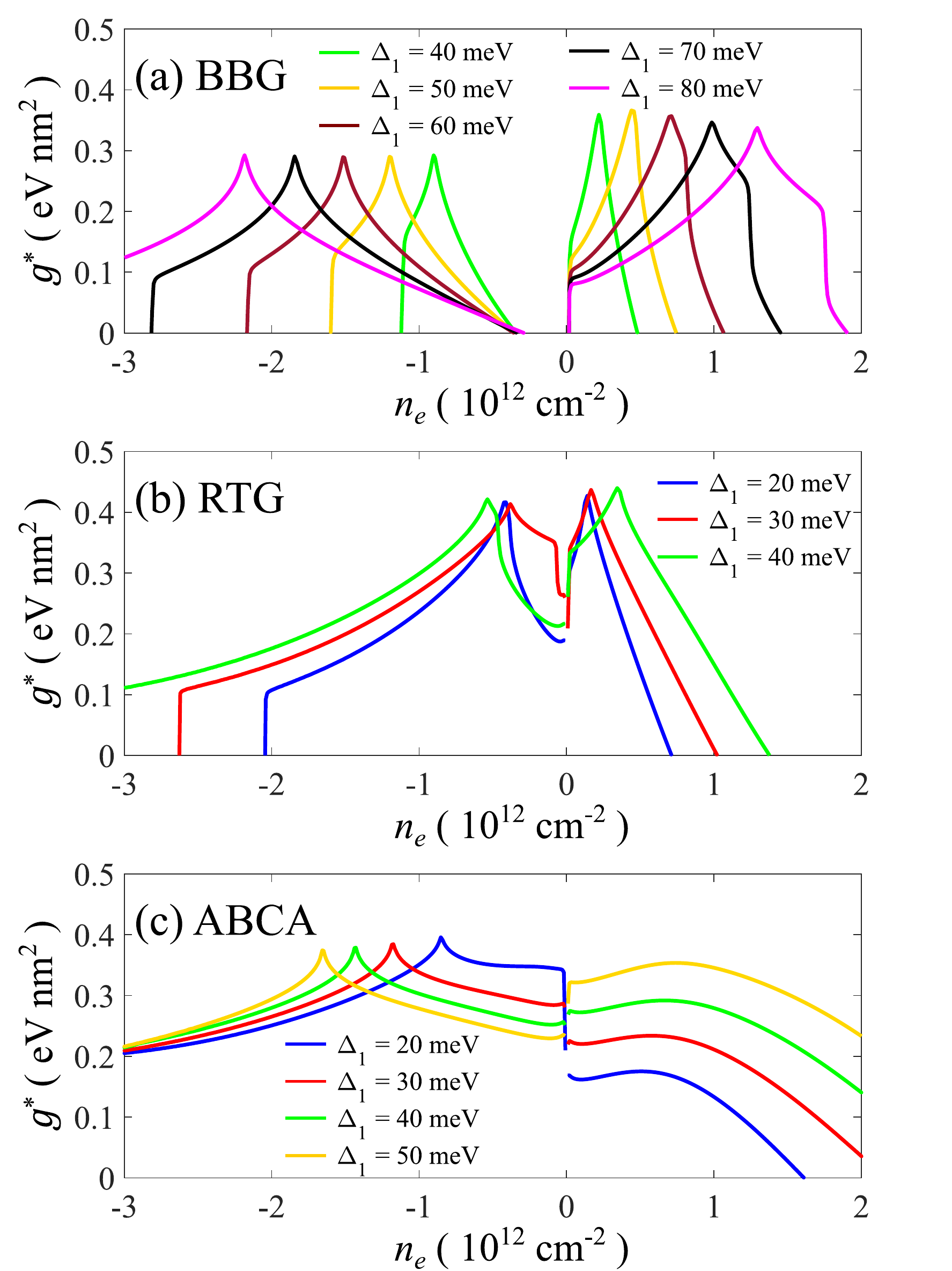}
	\caption{Effective attraction $g^*$ in unpolarized normal states. (a) BBG (b) RTG (c) ABCA tetralayer graphene. We choose energy cutoff $\Lambda =\min(2|\Delta_1|,100\text{meV})$ and $k_F$ is estimated by $\sqrt{4\pi |n_e|/f}$, where $f$ is the spin-valley degeneracy factor. $\Delta_1$ is a band parameter (defined in Appendix~\ref{App:Bands}) which can be tuned by a displacement field. 
	}
	\label{Fig:mu_star}
\end{figure}

\begin{figure}[t!]
	\includegraphics[width=0.4\textwidth]{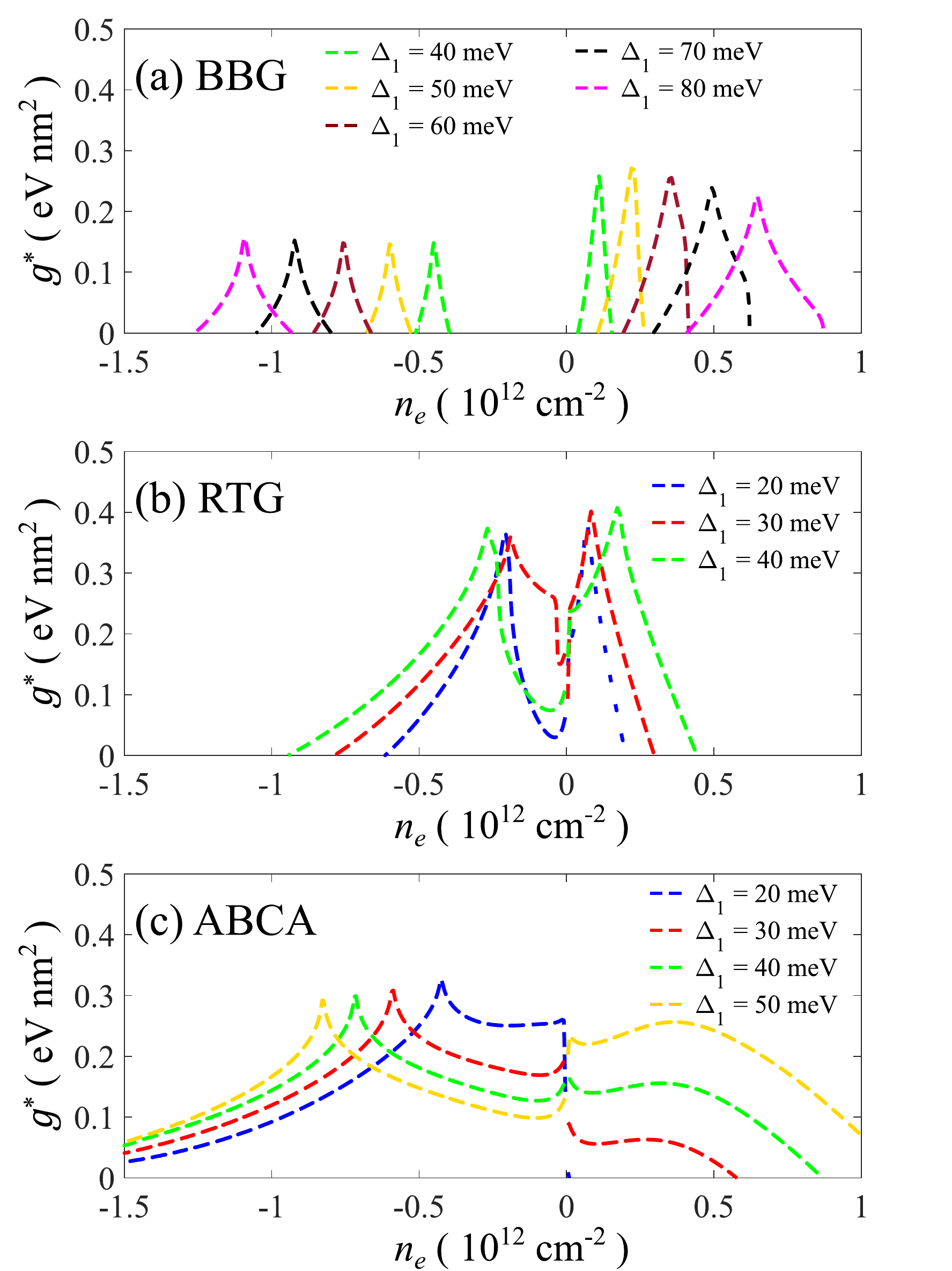}
	\caption{Effective attraction $g^*$ in spin-polarized normal states. (a) BBG (b) RTG (c) ABCA  tetralayer graphene. We choose energy cutoff $\Lambda =\min(2|\Delta_1|,100\text{meV})$ and $k_F$ is estimated by $\sqrt{4\pi |n_e|/f}$, where $f$ is the spin-valley degeneracy factor. $\Delta_1$ is a band parameter (defined in Appendix~\ref{App:Bands}) which can be tuned by a displacement field. 
	}
	\label{Fig:mu_star_SP}
\end{figure}

To achieve superconductivity, the phonon-mediated attraction must be stronger than the renormalized Coulomb repulsion in the low-energy regime so that effective Cooper pairing may occur, which then condense into the symmetry-broken superconducting BCS ground state. Equation~(\ref{Eq:U_R}) provides a quantitative estimate of the renormalized Coulomb repulsion within an energy window $[\mathcal{E}_F-\omega_c,\mathcal{E}_F+\omega_c]$. We can construct an effective BCS theory by replacing $g$ with the effective interaction $g^*=g-U_R(\mathcal{E}_F)$. We note that $g^*>0$ is a necessary condition for superconductivity, and the new gap equation is given by
\begin{align}
\label{Eq:LGE_mustar}\Delta_{s's}(\vex{k})=&\frac{1}{\mathcal{A}}\sum_{\vex{k}'}g_{\vex{k},\vex{k}'}^*\frac{\tanh\left[\frac{\mathcal{E}_{+b}(\vex{k}')-\mathcal{E}_F}{2k_BT}\right]}{2\mathcal{E}_{+b}(\vex{k}')-2\mathcal{E}_F}\Delta_{s's}(\vex{k}'),
\end{align}
where 
\begin{align}
\label{Eq:g_kk'_mustar}g_{\vex{k},\vex{k}'}^{*}=&g^*\sum_{\sigma,l}\left|\Phi_{+b;l\sigma}(\vex{k})\right|^2\left|\Phi_{+b;l\sigma}(\vex{k}')\right|^2.
\end{align}
In Eq.~(\ref{Eq:LGE_mustar}), we have ignored the explicit frequency dependence and mapped the frequency-dependent gap equations [Eq.~(\ref{Eq:LGE_freq})] to an effective BCS (frequency-independent) gap equation incorporating the $\mu^*$ effect. We emphasize the obvious fact that any superconductivity can only emerge if the effective interaction is attractive, i.e., $g > U_R(\mathcal{E}_F)$.  Also, $T_c$ would obviously depend on the relative strengths of the Coulomb repulsion $U_R(\mathcal{E}_F)$ and phonon-induced attraction $g$. The effective BCS approach here allows us to predict $T_c$ quantitatively without solving the extremely numerically demanding frequency-dependent Eliashberg equations. Strictly speaking, the Coulomb repulsion has a different form of matrix element after projecting to a single-band. However, the difference is negligible because of the layer-sublattice polarization in the low-energy bands of graphene multilayers. Thus, we stick to the current simplified approach, thus also avoiding possible large uncontrolled numerical errors in trying to solve the full frequency dependent self-consistent Eliashberg theory in a brute-force computational approach.

The value of $g^*$ depends on the ``isospin'' polarization in the normal states. We discuss the unpolarized (four-fold degenerate) normal states in Fig.~\ref{Fig:mu_star} and the spin-polarized (two-fold degenerate) normal states in Fig.~\ref{Fig:mu_star_SP}. The $g^*$ with unpolarized normal states is larger than $g^*$ with spin-polarized normal states because the Thomas-Fermi screening (intraband screening) crucially depends on DOS in the graphene multilayer. For BBG, $g^*$ is positive only in the vicinity of VHS, indicating that superconductivity is most likely found near VHS \cite{Chou2021_BBG}. For RTG and ABCA tetralayer graphene, we find that $g^*>0$ for a wide range of dopings, suggesting that superconductivity can prevail for a wide range of doping \cite{Chou2021_RTG_SC}. 

It is interesting to ask whether tuning gate distance ($2d$) or dielectric constant ($\epsilon$) can considerably modify $g^*$. For most doping $n_e$, $g^*$ is not sensitive to $d$ or $\epsilon$ because the large DOS strongly screens the Coulomb interaction (and, therefore, any additional screening by the gate and the background dielectric constant is quantitatively unimportant). For the regime where $g^*\le 0.2$eV$.$nm$^2$, we find that smaller $d$ (for $d<5$nm) and larger $\epsilon$ can considerably increase $g^*$, implying an enhancement in $T_c$. We note that $g^*$ is not sensitive to $d$ for $d>5$nm. Similar conclusions were reported previously for RTG \cite{Ghazaryan2021} and for BBG \cite{Chou2021_BBG}. In the next section, we show our calculated superconducting $T_c$ in various cases and discuss the interplay between phonon-mediated attraction and Coulomb repulsion.

\section{Numerical Results for superconducting $T_c$}\label{Sec:Results}

In this section, we present our numerical results for superconducting $T_c$ incorporating Coulomb repulsion. The $T_c$ is obtained by solving Eq.~(\ref{Eq:LGE_mustar}) numerically with a fine momentum mesh as discussed in Appendix~\ref{App:Numerics}. The results here are qualitatively consistent with the phonon-mediated-attraction-only results (i.e., the $\mu^*$ effect unimportant) for RTG and ABCA tetralayer graphene in Fig.~\ref{Fig:BCS}(b) and \ref{Fig:BCS}(c); the Coulomb repulsion significantly suppresses the superconducting region in BBG [Fig.~\ref{Fig:BCS}(a)] because the DOS is not significantly large in BBG, making screening less significant and hence producing a relatively large $\mu^*$. A thorough study for BBG has been done by us in Ref.~\cite{Chou2021_BBG}. In this section, we focus on RTG and ABCA tetralayer graphene, especially for the experimentally relevant regimes in RTG.

\subsection{Superconductivity from unpolarized normal states}

\begin{figure}[t!]
	\includegraphics[width=0.45\textwidth]{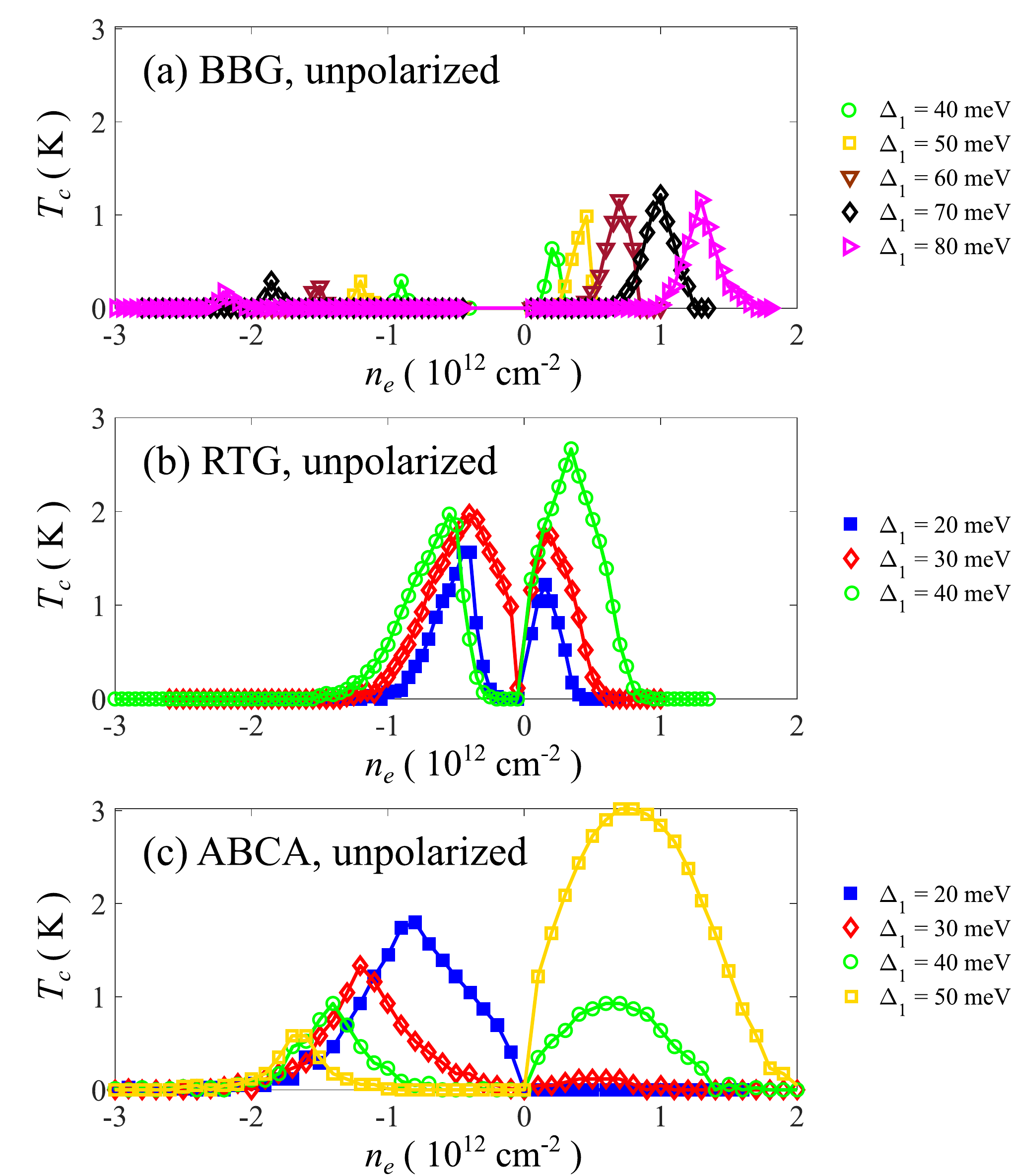}
	\caption{Superconducting $T_c$ for superconductivity incorporating Coulomb repulsion from unpolarized normal states. We use a dielectric constant $\epsilon=10$ and a gate distance parameter $d=20$nm for all the data. (a) BBG (b) RTG (c) ABCA tetralayer graphene. $\Delta_1$ is a band parameter (defined in Appendix~\ref{App:Bands}) which can be tuned by a displacement field. 
	}
	\label{Fig:Tc_e10_d20}
\end{figure}

We first discuss how the superconductivity arises from four-fold degenerate unpolarized normal states. As we discussed in the previous section, $g^*$ remains positive, indicating attractive interaction, for a wide range of doping in all three systems. In Fig.~\ref{Fig:Tc_e10_d20}, we plot $T_c$ as a function of $n_e$ with varied $\Delta_1$ for all three systems. Figs.~\ref{Fig:Tc_e10_d20}(b) and \ref{Fig:Tc_e10_d20}(c) show that observable superconductivity (say, $T_c> 20$mK) occurs in RTG and ABCA tetralayer graphene for a wide range of dopings, not just near VHS. However, this is not true for BBG as shown in Fig.~\ref{Fig:Tc_e10_d20}(a), where observable superconductivity exists only near VHS, and the highest $T_c$ is about $0.3$K ($1.2$K) for the hole (electron) doping. The very different results between BBG and other cases can be understood as arising from the quantitative difference in the DOS \cite{Chou2021_BBG}, which results in different $g^*$ in Fig.~\ref{Fig:mu_star} due to the quantitatively different $\mu^*$ effects.

\subsection{Superconductivity from spin-polarized normal states}

\begin{figure}[t!]
	\includegraphics[width=0.45\textwidth]{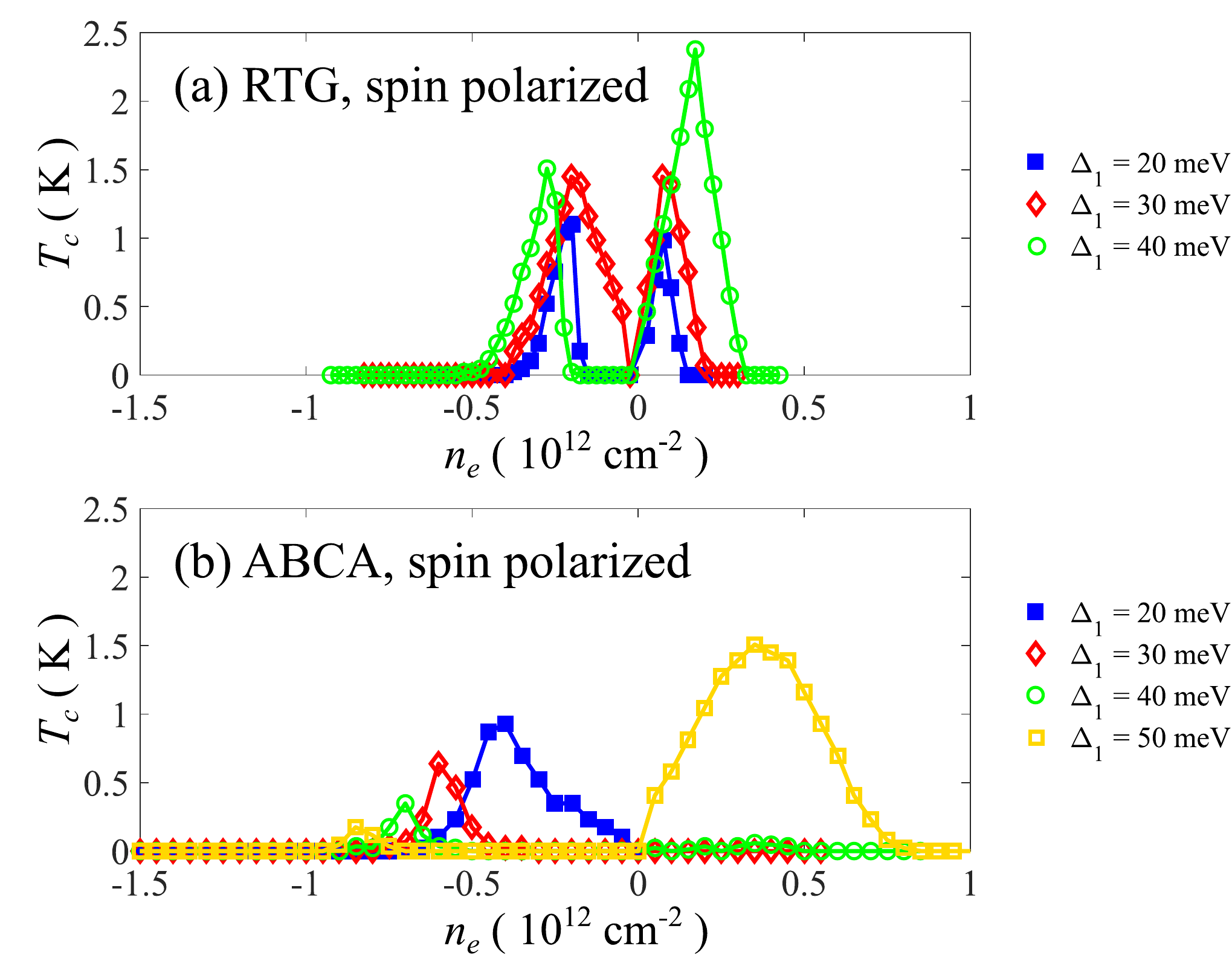}
	\caption{$T_c$ for superconductivity incorporating Coulomb repulsion. We use a dielectric constant $\epsilon=10$ and a gate distance parameter $d=20$nm for all the data.  (a) RTG (b) ABCA tetralayer graphene. $\Delta_1$ is a band parameter (defined in Appendix~\ref{App:Bands}) which can be tuned by a displacement field.  
	}
	\label{Fig:SP_Tc_e10_d20}
\end{figure}

For spin-polarized (two-fold degenerate) normal states, $\rho(\mathcal{E}_F)$ is half of the value of an unpolarized state at the same $\mathcal{E}_F$, so the intraband screening is weaker, resulting in a smaller $g^*$. We plot $T_c$ as a function of $n_e$ with varied $\Delta_1$ for RTG and ABCA tetralayer graphene in Fig.~\ref{Fig:SP_Tc_e10_d20}. (The superconducting $T_c$ for BBG, which is simply too small, is not quite visible with the same scale, and it was reported previously \cite{Chou2021_BBG}.) As expected, we find that $T_c$ is smaller in Fig.~\ref{Fig:SP_Tc_e10_d20} as compared to Fig.~\ref{Fig:Tc_e10_d20}(b) and \ref{Fig:Tc_e10_d20}(c), due to larger Coulomb repulsion because of weaker screening. Despite the reduction in $T_c$, observable superconductivity still prevails for a range of dopings, qualitatively similar to the unpolarized case. Again, this is quite different from spin-polarized normal states in BBG where any observable superconductivity is only expected near VHS \cite{Chou2021_BBG}. In Ref.~\cite{Chou2021_BBG}, the highest $T_c$ is around $20$mK ($0.5$K) in the hole (electron) doping.

\subsection{Superconductivity in RTG: Tuning Coulomb repulsion}\label{Sec:Tuning_Coulomb}

\begin{figure}[t!]
	\includegraphics[width=0.39\textwidth]{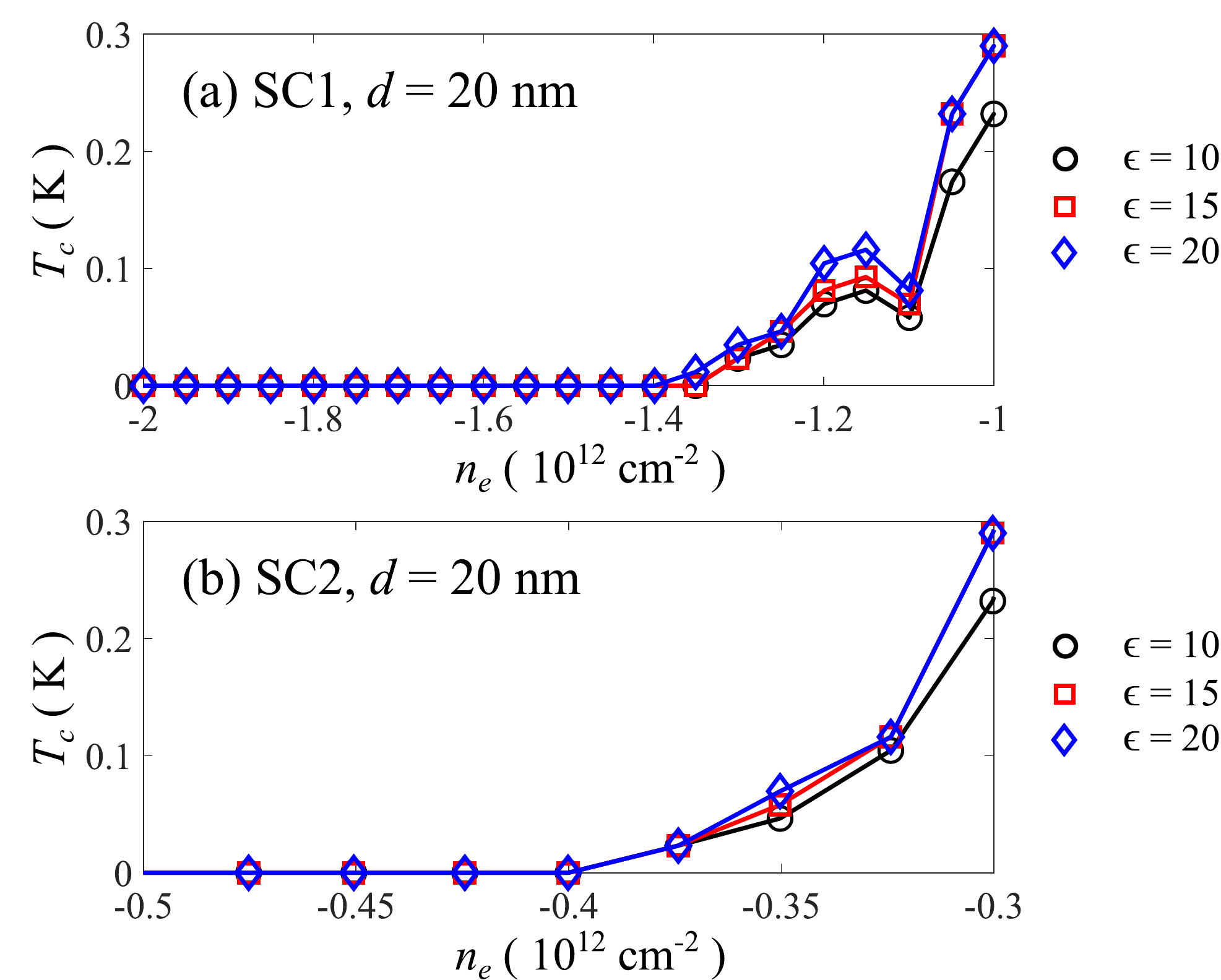}
	\caption{$T_c$ for RTG with different dielectric constant $\epsilon$. We use a gate distance parameter $d=20$nm for all the data. (a) SC1 regime corresponds to $\Delta_1=30$meV and unpolarized normal states. (b) SC2 regime corresponds to $\Delta_1=20$meV and spin-polarized normal states. $\Delta_1$ is a band parameter (defined in Appendix~\ref{App:Bands}) which can be tuned by a displacement field. 
	}
	\label{Fig:Tuning_e}
\end{figure}

\begin{figure}[t!]
	\includegraphics[width=0.4\textwidth]{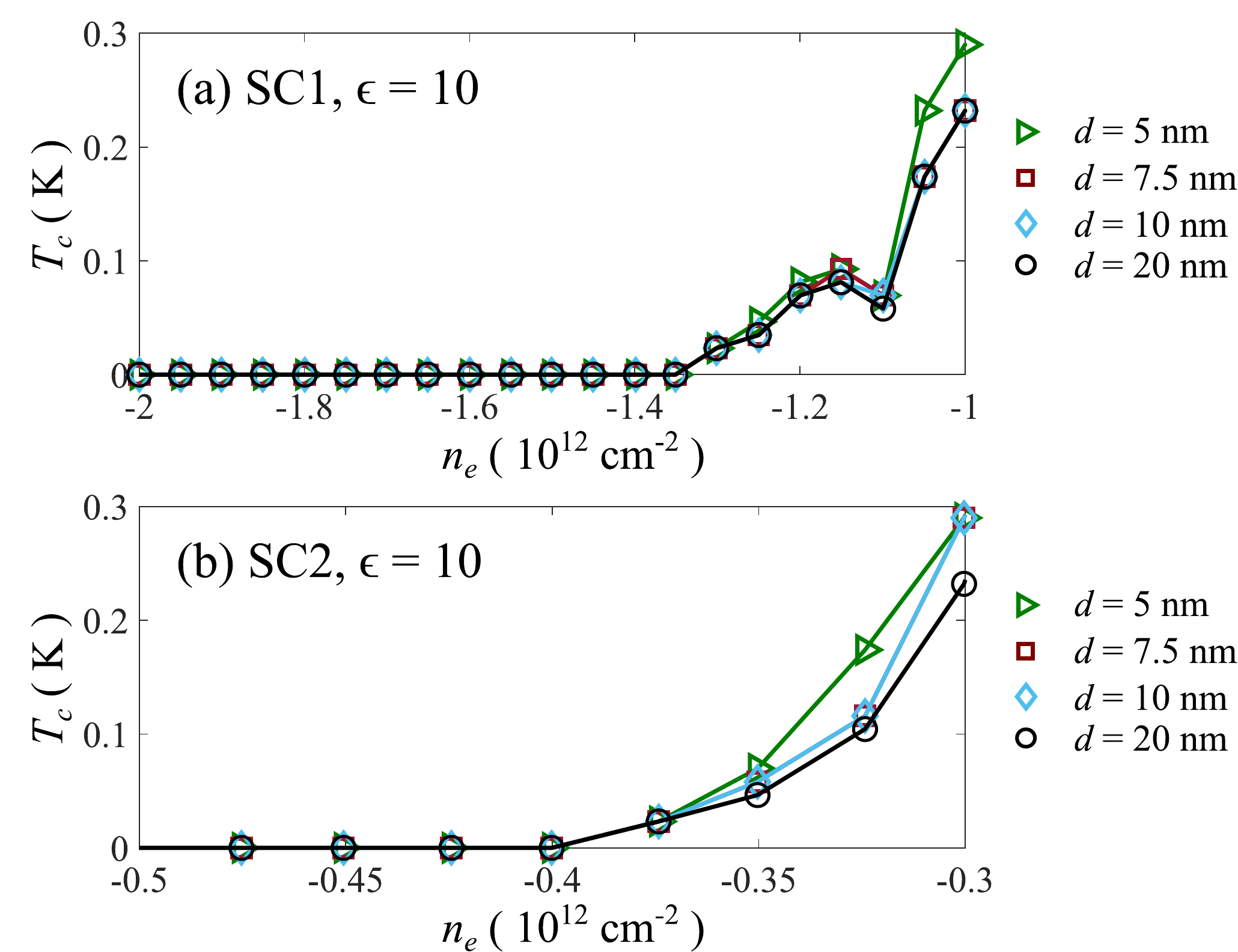}
	\caption{$T_c$ for RTG with different gate distance parameter $d$. We use a dielectric constant $\epsilon=10$ for all the data. (a) SC1 regime corresponds to $\Delta_1=30$meV and unpolarized normal states. (b) SC2 regime corresponds to $\Delta_1=20$meV and spin-polarized normal states. $\Delta_1$ is a band parameter (defined in Appendix~\ref{App:Bands}) which can be tuned by a displacement field. 
	}
	\label{Fig:Tuning_d}
\end{figure}

Based on the acoustic-phonon mechanism, suppressing Coulomb repulsion will boost $T_c$ because of the enhanced effective attraction near the Fermi surface. This can be achieved by decreasing the gate distance ($2d$) and increasing the effective dielectric constant ($\epsilon$). The main question is the amount of quantitative change in $T_c$. In our previous work on BBG \cite{Chou2021_BBG}, we provided the evolution of $T_c$ with different values of $d$ and $\epsilon$. Here, we focus on the regime where superconductivity is observed in RTG \cite{Zhou2021_SC_RTG}. 

In the RTG experiment, both spin-singlet (coined SC1) and non-spin-singlet (coined SC2) superconducting states were observed \cite{Zhou2021_SC_RTG}.
We assume $\Delta_1=30$ meV ($\Delta_1$ corresponds to the displacement field), $n_e\approx -1.9\times 10^{12}$cm$^{-2}$, and unpolarized normal states for SC1; we assume $\Delta_1=20$ meV, $n_e\approx -0.5\times 10^{12}$cm$^{-2}$, and spin-polarized normal states for the SC2 regime. First of all, we do not find observable $T_c$ at $n_e\approx -1.9\times 10^{12}$cm$^{-2}$ for SC1 or $n_e\approx -0.5\times 10^{12}$cm$^{-2}$ for SC2. This is likely a quantitative issue due to parameters used in our theory (such as $g$, band parameters, $n_e$, etc). Therefore, we investigate the regimes with observable $T_c$ close to SC1 and SC2 dopings. In Fig.~\ref{Fig:Tuning_e}, $T_c$ with a few representative dielectric constants ($\epsilon$) is plotted. Larger $\epsilon$ indeed enhances $T_c$, but the enhancement is not significant for states near SC1 or SC2 regime. In Fig.~\ref{Fig:Tuning_d}, we vary the gate distance ($2d$) and plot the corresponding $T_c$. $T_c$ gets larger for a smaller $d$, but the enhancement is not outstanding for $d>5$nm, consistent with our finding in the $g^*$ previously. Note that the $T_c$ remains essentially independent of $\epsilon$ or $d$ for regimes with $T_c>0.5$K. This can be understood by the associated large DOS in those regimes, where Coulomb repulsion is strongly screened by the graphene carriers themselves, essentially independent of $\epsilon$ or $d$.

\subsection{Superconductivity in RTG: Varying $g$}\label{Sec:New_G}

\begin{figure}[t!]
	\includegraphics[width=0.375\textwidth]{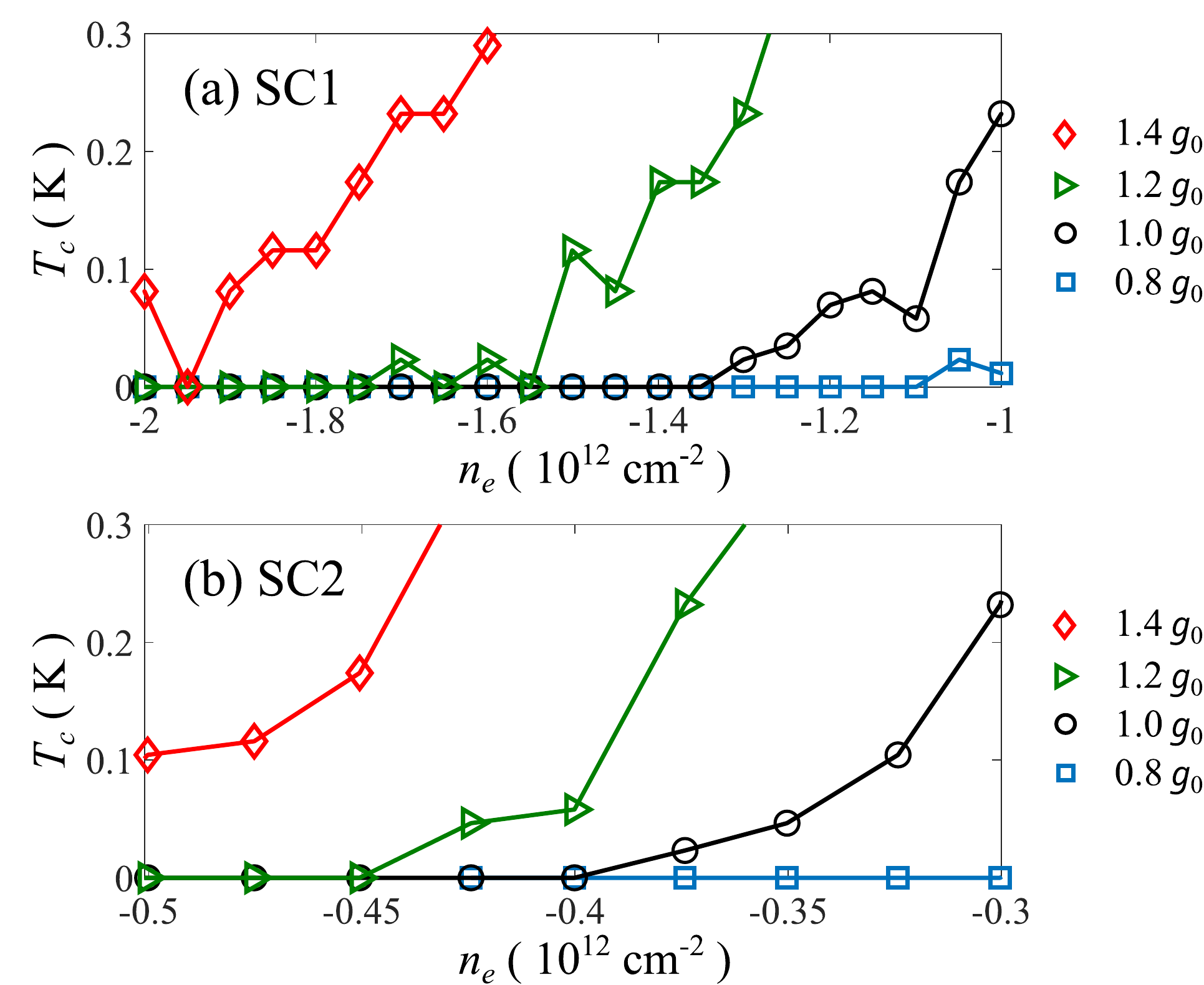}
	\caption{Superconductivity with different values of phonon-mediated attraction $g$. We use a dielectric constant $\epsilon=10$ and a gate distance parameter $d=20$nm for all the data, and $g_0=474$ meV$\cdot$nm$^2$. (a) SC1 regime corresponds to $\Delta_1=30$meV and assume unpolarized normal states. (b) SC2 regime corresponds to $\Delta_1=20$meV and assume spin-polarized normal states. $\Delta_1$ is a band parameter (defined in Appendix~\ref{App:Bands}) which can be tuned by a displacement field. 
	}
	\label{Fig:New_G}
\end{figure}

Based on our theory with $g=g_0=474$ meV$\cdot$nm$^2$, we cannot reproduce observable $T_c$ in SC1 ($n_e\approx -1.9\times 10^{12}$cm$^{-2}$) or SC2 regime ($n_e\approx -0.5\times 10^{12}$cm$^{-2}$). This is a quantitative issue because the value of $g$ is not precisely known since the deformation potential coupling is often unknown \cite{Hwang2008,Wu2019_phonon}, and the $T_c$ is quite sensitive to $g$. To investigate this issue, we vary the value of $g$ and plot the corresponding $T_c$ in Fig.~\ref{Fig:New_G}. We find that comparable $T_c$ (to the experiment \cite{Zhou2021_SC_RTG}) can be reproduced with an enhanced value of phonon-mediated attraction $1.4g_0$. Since $g=D^2/(\rho_m v_s^2)$, a slightly larger $D$ and/or a slightly smaller $v_s$ can result in a larger $g$. An interesting finding here is that our predicted $T_c$ for SC1 is slightly smaller than the $T_c$ for SC2, while it is found that SC1 is stronger than SC2 in the RTG experiment \cite{Zhou2021_SC_RTG}. We discuss possible explanations in Sec.~\ref{Sec:Discussion}.

\section{Discussion}\label{Sec:Discussion}

We study acoustic-phonon-mediated superconductivity in untwisted graphene multilayers-- BBG, RTG, and ABCA tetralayer graphene including effects of direct Coulomb repulsion. The $SU(2)\times SU(2)$ symmetric acoustic-phonon-mediated attraction naturally favors intrasublattice pairings in untwisted graphene multilayers, making $s$-wave spin-singlet and $f$-wave spin-triplet pairings dominant and degenerate. We develop a simplified, but quantitatively predictive, theory incorporating both phonon-mediated attraction and direct Coulomb repulsion. Within the mean-field approximation, we reproduce the recently experimentally-observed superconductivity phenomenology in BBG and RTG, and we further predict the existence of superconductivity in ABCA tetralayer graphene, which should be experimentally investigated. Our theory captures the qualitative and semi-quantitative features of the experiments \cite{Zhou2021_SC_RTG,Zhou2021_BBG}, suggesting that superconductivity in graphene untwisted multilayers is likely due to acoustic phonons.

To understand why and how the acoustic-phonon mechanism can explain the BBG \cite{Zhou2021_BBG} and RTG \cite{Zhou2021_SC_RTG} experiments, one has to take into account the Coulomb repulsion which causes a suppression of the predicted $T_c$ leading to agreement with experiments. Because the BBG band generates a smaller DOS resulting in weaker screening, the Coulomb repulsion suppresses superconductivity for most doping densities except near VHS. On the other hand, the large DOS of RTG efficiently screens Coulomb repulsion and results in observable superconducting states for a wide range of doping. Thus, our results provide natural explanations to the BBG and RTG experiments without any fine-tuning or arbitrary data fitting, as we explain in the following. 

In the BBG experiment, a sufficiently large in-plane magnetic field, which suppresses a competing order, is required to observe superconductivity \cite{Zhou2021_BBG}. Based on our theory, observable superconductivity (i.e., $T>20$mK) can happen only near VHS. The applied in-plane magnetic field likely suppresses the competing order, which, if presents, preempts superconductivity, and spin-triplet superconductivity manifests itself in the absence of the competing order. 

In the RTG experiment, superconductivity is observed away from VHS without a magnetic field, and spin-singlet (spin-triplet) superconductivity emerges from unpolarzied normal states (spin-polarized normal states) \cite{Zhou2021_SC_RTG}. Our theory can naturally explain the results because the $SU(2)\times SU(2)$ symmetry in acoustic-phonon-mediated attraction favors $s$-wave spin-singlet and $f$-wave spin-triplet pairings \cite{Wu2019_phonon,Chou2021_RTG_SC}. The $s$-wave spin-singlet is usually the dominating pairing for unpolarized normal states (four-fold degenerate) because the subleading pairings (e.g., optical phonons \cite{Wu2018}) can enhance $s$-wave spin-singlet pairings also. For spin-polarized normal states (two-fold degenerate), spin-singlet pairings are suppressed, and $f$-wave spin-triplet pairing becomes the leading pairing instability. In RTG, the absence of superconductivity near VHS or in the regime with large DOS is due to the competing correlation-induced instabilities from interaction. Note that the Stoner-type instability is very sensitive to the value of DOS at $\mathcal{E}_F$, but this is not true for the superconducting instability \cite{Lothman2017}. As a result, observable superconducting states can be found away from VHS in the RTG experiment. Based on our results, both the SC1 and SC2 doping densities correspond to the tail regions of observable superconductivity, implying that the majority of superconductivity is superseded by correlation-induced instabilities (partially isospin polarized states \cite{Zhou2021_SC_RTG}) except for narrow regions of SC1 and SC2.

An interesting question is whether the predicted superconductivity is robust against disorder or scatterings introduced at the sample boundary (e.g., Refs.~\cite{Wakabayashi2009electronic} and \cite{Walter2018}). While intervalley scattering can be suppressed in clean devices near perfect charge neutrality \cite{Akhmerov2008}, charge impurities at edges can cause intervalley scattering \cite{Wakabayashi2009electronic}.
The $s$-wave spin-singlet superconductivity is robust against weak charge (but not magnetic) disorder as showed by the Anderson theorem. However, the $f$-wave spin-polarized spin-triplet (valley-singlet) pairing, which we predict for SC2 in RTG and superconductivity in BBG, is fragile in the presence of intervalley scatterings. This is mathematically analogous to the suppression of $s$-wave spin-singlet pairing due to spin-flipping scatterings \cite{Fulde1966}. To examine the role of intervalley scattering, we first estimate the coherence lengths of SC2 in Ref.~\cite{Zhou2021_SC_RTG}. We obtain the BCS coherence length $\xi_{\text{BCS}}=\frac{\hbar v_F}{\pi \Delta_0}\approx 1.38\mu$m (using $T_c=50$mK and $v_F=5\times10^4$m/s) and Ginzburg-Landau coherence length $\xi_{\text{GL}}=\sqrt{\frac{h/(2e)}{2\pi B_{c,\perp}}}\approx0.57\mu$m (using $B_{c,\perp}=1$mT). The distance between nearby contacts is around $2\mu$m, which is not significantly larger than the coherence lengths, suggesting that scatterings at boundary might affect the superconductivity. Assuming an intervalley scattering time $\tau_s$ and following Ref.~\cite{Sau2012}, we obtain an expression of the pairing potential strength ($\tilde\Delta$) perturbed by intervalley scatterings (at the second order) as follows (see Appendix~\ref{App:Intervalley} for derivations):
\begin{align}\label{Eq:Intervalley}
|\tilde\Delta(\tau_s,\Delta)|=|\Delta|\left[1-\frac{1}{2^{1/3}}\left(\frac{\mathcal{C}}{|\Delta|\tau_s}\right)^{2/3}\right],
\end{align}
where $\Delta$ is the pairing potential without intervalley scatterings and $\mathcal{C}$ is a constant encoding the average over Fermi surface and DOS at $\mathcal{E}_F$. Equation~(\ref{Eq:Intervalley}) describes the pair breaking effect due to weak intervalley scatterings in spin-polarized spin-triplet superconductivity. Assuming that the intervalley scattering is strong at the edges, the intervalley scattering rate is limited by the sample size ($L$), i.e. $\tau_s\sim L/v_F$. This results in the quantity $|\Delta|\tau_s\sim L/\xi_{\text{BCS}}$. Thus the superconductivity can survive for devices that are larger than the coherence length. This can also be understood as superconductivity in the presence of pair breaking edge disorder, where the superconducting order parameter goes to zero at the edges. The superconductivity is expected to revive on the scale of $\xi$, which can occur if the system is larger than the coherence length. The results here also qualitatively apply to superconductivity in BBG.

Since the pairing glue comes from phonons in our theory, suppressing Coulomb repulsion (i.e., increasing $\epsilon$ or decreasing $d$) generically enhances superconducting $T_c$. As we discussed in Secs.~\ref{Sec:g_star} and \ref{Sec:Tuning_Coulomb}, changing $\epsilon$ and $d$ might not result in significantly different $T_c$ in RTG because the Coulomb repulsion is in the strong screening regime. In particular, the gate distance dependence is quite weak for $d>5$nm. (A similar conclusion was drawn in Ref.~\cite{Ghazaryan2021}.) For BBG, $T_c$ is more sensitive to the value of $\epsilon$ and $d$ as we pointed out previously in Ref.~\cite{Chou2021_BBG}. This can be understood by the smaller DOS in BBG, such that the intraband screening is not fully suppressing the dependence of $d$ or $\epsilon$. The possible enhancement of $T_c$ by reducing gate/dielectric screening is a testable theoretical prediction of our theory.

Another important question is whether the electron-phonon coupling constant is correctly estimated in our theory as the deformation potential $D$ is not precisely known \cite{Hwang2008,Wu2019_phonon}. In Sec.~\ref{Sec:New_G}, we find that $g=1.4g_0$ can reproduce the comparable $T_c$ for SC1 and SC2 in RTG. In the RTG experiments, $T_{\text{BKT}}\approx 100$mK for SC1 and $T_{\text{BKT}}< 50$mK for SC2 were reported \cite{Zhou2021_SC_RTG}. However, our theory with $g=1.4g_0$ gives $T_c\approx80$mK for SC1 and $T_c\approx 100$mK for SC2. This raises a puzzle as our predicted $T_c$'s are in the opposite order of the experimental results. The discrepancy might be understood by the fragile nature of non-spin-singlet superconductivity (SC2), which can be suppressed easily by disorder or intervalley scatterings (e.g., scattering at sample boundary) in the experiments. Another possible explanation is that there exists a subleading pairing mechanism (such as optical phonons \cite{Wu2018} or other interaction-induced pairings \cite{Chatterjee2021,Ghazaryan2021,Dong2021,Cea2021,Szabo2021,You2021}) that contributes to SC1 but not SC2. Regardless of the possible explanation, the acoustic-phonon-mediated pairing is still the dominating gluing mechanism. We leave the puzzling discrepancy as an open question for future studies, which also requires the availability of more experimental results.

Now, we discuss a number of predictions based on the acoustic-phonon theory. An interesting prediction based on our theory is that a sufficient large in-plane magnetic field can destroy the $s$-wave spin-singlet pairing, and then the $f$-wave equal-spin pairing becomes the leading superconducting instability \cite{Chou2021_RTG_SC}. In addition, it is possible that an applied in-plane magnetic field can induce new superconducting phases in RTG by suppressing the competing ordered states, similar to the field-induced superconductivity in BBG. Our theory predicts observable superconductivity not just for hole doping but also for electron doping for BBG, RTG, and ABCA tetralayer graphene, and we find that a larger $|\Delta_1|$ generally enhances the superconducting $T_c$ for electron doping. 

One important consequence of strong electron-acoustic phonon coupling is that the finite-temperature resistivity should develop a linear-in-$T$ resistivity for $T>T_{\text{onset}}$ and a $T^4$ resistivity for $T<T_{\text{onset}}$ \cite{Hwang2008}. We estimate that $T_{\text{onset}}$ is above 10K-20K \cite{Hwang2008,Min2011} for both BBG and RTG. The electron-phonon coupling parameter extracted from such a linear-in-$T$ resistivity should have approximate consistency with the observed $T_c$ \cite{Min2011,Wu2019_phonon,Li2020,Polshyn2019,Cao2020PRL,Chu2021phonons}. The same is true for spin or valley fluctuation mediated superconductivity, too. In the RTG experiment \cite{Zhou2021_SC_RTG} (BBG experiment \cite{Zhou2021_BBG}), a linear-in-$T$ resistivity is not seen for $T\le 20$K ($T\le 1.5$K), where the highest temperature appears to be smaller than our estimated $T_{\text{onset}}>20$K. Again, based on our theory, there should be a phonon-induced linear-in-$T$ resistivity for $T>20$K above the superconducting state. This should be investigated experimentally by extending the conductivity measurements to $T=10$K$-50$K regime.

Finally, we comment on the universal theory of superconductivity in graphene-based materials (including twisted and untwisted materials). It is likely that the electron-acoustic-phonon mechanism accounts for superconductivity in all graphene-based materials provided that acoustic phonons can explain the distinct superconductivity phenomenology in BBG \cite{Zhou2021_BBG}, in RTG \cite{Zhou2021_SC_RTG}, and in twisted bilayer graphene \cite{Wu2019_phonon}. In addition, several experiments on magic-angle moir\'e graphenes show that superconductivity is more robust \cite{Lu2019,Saito2020independent,Stepanov2020untying,Liu2021tuning}, i.e., it can exist without any nearby correlated insulating states, hence arguing against a correlation-induced mechanism. Thus, it is natural to suspect that superconductivity and correlated states most likely come from different origins \cite{Chou2019,Wu2018,Lian2019,Wu2019_phonon,Lewandowski2021,Shavit2021}, and the acoustic-phonon mechanism can explain the superconductivity phenomenology. We emphasize that the interactions are still essential as they can induce competing orders, suppressing and preempting superconductivity. Our qualitative picture is that all graphene superconductivity is induced by acoustic phonons, but competing strongly correlated non-superconducting phases arising from electron-electron interactions may arise, competing with and occasionally suppressing the superconducting phase. In summary, we present a systematic theory, incorporating electron-phonon couplings and Coulomb repulsion, for superconductivity in untwisted graphene multilayers. We obtain $T_c$ in reasonable agreement with the experimental observation, and we speculate that acoustic phonons are responsible for all superconductivity in graphene-based materials in general.

\begin{acknowledgments}
We thank Anton Akhmerov and Matt Foster for useful discussions. This work is supported by the Laboratory for Physical Sciences (Y.-Z.C. and S.D.S.), by JQI-NSF-PFC (Y.-Z.C.), and by NSF DMR1555135 (J.D.S.). F.W. is supported by National Key R$\&$D Program of China 2021YFA1401300 and start-up funding of Wuhan University.
\end{acknowledgments}	
\appendix

\section{Band structure}\label{App:Bands}

\subsection{BBG band structure}

We adopt the $\vex{k}\cdot\vex{p}$ Hamiltonian from Ref.~\cite{Jung2014}. The $\hat{h}_{\tau}(\vex{k})$ in the main text is given by
\begin{align}
	\hat{h}_{\tau}^{(2)}(\vex{k})
	=\left[\begin{array}{cccc}
		-\Delta_1 & v_0\Pi^{\dagger}(\vex{k}) & -v_4\Pi^{\dagger}(\vex{k}) & -v_3\Pi(\vex{k}) \\[2mm]
		v_0\Pi(\vex{k}) & \Delta'-\Delta_1 & t_1 & -v_4\Pi^{\dagger}(\vex{k}) \\[2mm]
		-v_4\Pi(\vex{k})& t_1 & \Delta'+\Delta_1 & v_0\Pi^{\dagger}(\vex{k}) \\[2mm]
		-v_3\Pi^{\dagger}(\vex{k})& -v_4\Pi(\vex{k}) & v_0\Pi(\vex{k})  & \Delta_1
	\end{array}
	\right],
\end{align}
where $\Pi(\vex{k})=\tau k_x+ik_y$, $a_0$ is the lattice constant of graphene, and $\Delta_1$ encodes the electric potential difference from the displacement field. The other parameters are given by $v_0/a_0=2.261$ eV, $v_3/a_0=0.245$ eV, $v_4/a_0=0.12$ eV, $t_1=0.361$ eV, and $\Delta'=0.015$ eV. The basis of the matrix is (1A,1B,2A,2B).

\subsection{RTG band structure}

The 6-by-6 matrix $\hat{h}_{\tau}(\vex{k})$ is given by \cite{Zhang2010,Zhou2021}
\begin{widetext}
	\begin{align}\label{Eq:h_k}
		\hat{h}_{\tau}^{(3)}(\vex{k})=\left[\begin{array}{cccccc}
			\Delta_1+\Delta_2+\delta& \frac{1}{2}\gamma_2 & v_0\Pi^{\dagger}_{\vex{k}} & v_4\Pi^{\dagger}_{\vex{k}} & v_3\Pi_{\vex{k}} & 0 \\[2mm]
			\frac{1}{2}\gamma_2 & \Delta_2-\Delta_1+\delta & 0 & v_3\Pi^{\dagger}_{\vex{k}} & v_4\Pi_{\vex{k}} & v_0\Pi_{\vex{k}} \\[2mm]
			v_0\Pi_{\vex{k}} & 0 & \Delta_1+\Delta_2 & \gamma_1 & v_4\Pi^{\dagger}_{\vex{k}} & 0 \\[2mm]
			v_4\Pi_{\vex{k}} & v_3\Pi_{\vex{k}} & \gamma_1 & -2\Delta_2 & v_0\Pi^{\dagger}_{\vex{k}} & v_4\Pi^{\dagger}_{\vex{k}} \\[2mm]
			v_3\Pi^{\dagger}_{\vex{k}} & v_4\Pi^{\dagger}_{\vex{k}} & v_4\Pi_{\vex{k}} & v_0\Pi_{\vex{k}} & -2\Delta_2 & \gamma_1 \\[2mm]
			0 & v_0\Pi^{\dagger}_{\vex{k}} & 0 & v_4\Pi_{\vex{k}} & \gamma_1 & \Delta_2-\Delta_1
		\end{array}
		\right],
	\end{align}
\end{widetext}
where $\Pi_{\vex{k}}=\tau k_x+ik_y$ ($\tau= 1,-1$ for valleys K and $-$K respectively), $v_j=\frac{\sqrt{3}}{2}\gamma_ja_0$, $\gamma_j$ is the bare hopping matrix element, and $a_0=0.246$nm is the lattice constant of graphene. The basis of $\hat{h}_{\tau}(\vex{k})$ is (1A,3B,1B,2A,2B,3A). Note that the first two elements, 1A and 3B, are the low-energy sites as discussed in the main text.\\

We use the same parameters in Ref.~\cite{Zhou2021}. Specifically, $\gamma_0=3.1$eV, $\gamma_1=0.38$eV, $\gamma_2=-0.015$eV, $\gamma_3=-0.29$eV, $\gamma_4=-0.141$eV, $\delta=-0.0105$eV, and $\Delta_2=-0.0023$eV. The value of $\Delta_1$ corresponds to the out-of-plane displacement field, and we vary it from 10 to 40meV.

\subsection{ABCA band structure}

Building on the $\vex{k}\cdot\vex{p}$ band model for RTG \cite{Zhang2010,Zhou2021}, we obtain a $\vex{k}\cdot\vex{p}$ band model for ABCA-stacked tetralayer graphene, given by
\begin{widetext}
	\begin{align}
		\hat{h}^{(4)}_{\tau}(\vex{k})=\left[
		\begin{array}{cccccccc}
			\delta-\Delta_1+\Delta_2 & v_0 \Pi_{\vex{k}}^{\dagger} & v_4 \Pi_{\vex{k}}^{\dagger} & v_3 \Pi_{\vex{k}} & 0 & \frac{1}{2}\gamma_2 & 0 & 0 \\
			v_0 \Pi_{\vex{k}} & \Delta_2-\Delta_1 & \gamma_1 & v_4 \Pi_{\vex{k}}^{\dagger} & 0 & 0 & 0 & 0 \\
			v_4 \Pi_{\vex{k}} & \gamma_1 & -\frac{\Delta_1}{3}-2 \Delta_2 & v_0 \Pi_{\vex{k}}^{\dagger} & v_4 \Pi_{\vex{k}}^{\dagger} & v_3 \Pi_{\vex{k}} & 0 & \frac{1}{2}\gamma_2 \\
			v_3 \Pi_{\vex{k}}^{\dagger} & v_4 \Pi_{\vex{k}} & v_0 \Pi_{\vex{k}} & -\frac{\Delta_1}{3}-2 \Delta_2+\Delta_3 & \gamma_1 & v_4 \Pi_{\vex{k}}^{\dagger} & 0 & 0 \\
			0 & 0 & v_4 \Pi_{\vex{k}} & \gamma_1 & \frac{\Delta_1}{3}-2 \Delta_2+\Delta_3 & v_0 \Pi_{\vex{k}}^{\dagger} & v_4 \Pi_{\vex{k}}^{\dagger} & v_3 \Pi_{\vex{k}} \\
			\frac{1}{2}\gamma_2 & 0 & v_3 \Pi_{\vex{k}}^{\dagger} & v_4 \Pi_{\vex{k}} & v_0 \Pi_{\vex{k}} & \frac{\Delta_1}{3}-2 \Delta_2 & \gamma_1 & v_4 \Pi_{\vex{k}}^{\dagger} \\
			0 & 0 & 0 & 0 & v_4 \Pi_{\vex{k}} & \gamma_1 & \Delta_1+\Delta_2 & v_0 \Pi_{\vex{k}}^{\dagger} \\
			0 & 0 & \frac{1}{2}\gamma_2 & 0 & v_3 \Pi_{\vex{k}}^{\dagger} & v_4 \Pi_{\vex{k}} & v_0 \Pi_{\vex{k}} & \delta+\Delta_1+\Delta_2 \\
		\end{array}
		\right].
	\end{align}
\end{widetext}
The parameters in $\hat{h}^{(4)}_{\tau}(\vex{k})$ are the same as the band parameters in RTG. The basis of the matrix is (1A,1B,2A,2B,3A,3B,4A,4B).

The hole-doping low-energy band is qualitatively similar to RTG while the electron-doping low-energy band is distinct from RTG or BBG, i.e., it manifest only one Fermi pocket (instead of three) for a given valley and spin. The $\vex{k}\cdot\vex{p}$ band structure here is qualitatively similar to the rhombohedral graphite system \cite{Shi2020}.

\section{Numerical procedures}\label{App:Numerics}

The DOS profiles in Fig.~\ref{Fig:DOS} are constructed with $10^4\times 10^4$ momentum mesh with a momentum spacing $\Delta k \approx 2\times 10^{-5}a_0^{-1}$. We use this momentum mesh to construct a map between $\mathcal{E}_F$ and $n_e$ for all the numerical results. This causes some quantitative difference between Fig.~\ref{Fig:BCS}(b) and Ref.~\cite{Chou2021_RTG_SC}, where $n_e$ is determined by a much smaller momentum mesh. 

The linearized gap equation can be viewed as an eigenvalue problem. The goal is to find the highest temperature such that Eq.~(\ref{Eq:LGE_1}) or (\ref{Eq:LGE_mustar}) is satisfied. To evaluate this numerically, we consider a fine momentum mesh with $\Delta k \approx0.002 a_0^{-1}$ and keep 5000 low-energy states. We have tested finer momentum meshes with more states kept, and the results are essentially the same, suggesting convergence.

In Eq.~(\ref{Eq:LGE_mustar}), DOS is needed for estimating $g^*$. We use the DOS profiles in Fig.~\ref{Fig:DOS} to construct a map between $\mathcal{E}_F$ and DOS. Note that the momentum mesh used in the calculations of Fig.~\ref{Fig:DOS} is much finer than the momentum mesh for extracting $T_c$.

\section{Eliashberg theory}\label{App:Elia_Th}

In this appendix, we present a derivation of the Eliashberg theory with a path integral approach \cite{Protter2021}. In the low-temperature low-doping limit, we focus only on one of the low-energy bands. The imaginary-time action with projection to band $b$ ($b$ index is sometimes dropped for simplicity) is given by $\mathcal{S}=\mathcal{S}_0+\mathcal{S}_I$, where
\begin{align}
	\mathcal{S}_0=&\sum_{\tau,s}\sum_{k}\bar{c}_{\tau s, k}\left[-i\omega_n+\mathcal{E}_{\tau}(\vex{k})-\mathcal{E}_F\right]c_{\tau s,k},\\
	\nonumber\mathcal{S}_I\approx&\frac{1}{2\beta\mathcal{A}}\sum_{\tau,s}\sum_{k,k'}W(k,k')\bar{c}_{\tau s,k}c_{\tau s,k}\bar{c}_{\tau s,k'}c_{\tau s,k'}\\
	&-\!\frac{1}{\beta\mathcal{A}}\sum_{s,s'}\sum_{k,k'}W(k,k')\bar{c}_{+s,k}\bar{c}_{-s',-k}c_{-s',-l'}c_{+s,k'}.
\end{align}
In the above expressions, $W_{k,k'}=V^{(b)}_g(k,k')-V^{(b)}_{\text{TF}}(k,k')$,
\begin{align}
\nonumber V^{(b)}_{g}(k,k')=&V_g(\omega_n-\omega_n',\vex{k}-\vex{k}')\\
&\times\sum_{\sigma,l}\left|\Phi_{+,b;l,\sigma}(\vex{k})\right|^2\left|\Phi_{+,b;l,\sigma}(\vex{k}')\right|^2,\\
V^{(b)}_{\text{TF}}(k,k')=&V_{\text{TF}}(\vex{k}-\vex{k}')\left|\sum_{\sigma,l}\Phi^*_{+,b;l,\sigma}(\vex{k})\Phi_{+,b;l,\sigma}(\vex{k}')\right|^2.
\end{align}
In $\mathcal{S}_I$, only the intra-valley $u$ channel and the inter-valley BCS channel are included. We note that the band projection matrix elements are the same for the intra-valley $u$-channel and the inter-valley BCS channel. This choice of interaction terms allows us to derive Eliashberg equations straightforwardly. 

To decouple interactions, we introduce Hubbard-Stratonovich field, and $\mathcal{S}_I$ becomes
\begin{align}
	\nonumber\mathcal{S}_I\rightarrow&\sum_{s,s'}\sum_{k}\left[\bar{\Delta}_{ss'}(k)c_{-,s',-k}c_{+,s,k}
	+\bar{c}_{+,s,k}\bar{c}_{-,s',-k}\Delta_{ss'}(k)\right]\\
	\nonumber&+\sum_{\tau}\sum_s\sum_{k}\left[i\Xi_{\tau s}(k)\bar{c}_{\tau,s,k}c_{\tau,s,k}\right]\\
	\nonumber&+ \mathcal{\beta}\mathcal{A}\sum_{s,s'}\sum_{k,k'}\bar{\Delta}_{ss'}(k)W^{-1}(k,k')\Delta_{ss'}(k')\\
	&+ \frac{\mathcal{\beta}\mathcal{A}}{2}\sum_{\tau}\sum_{s}\sum_{k,k'}\Xi_{\tau s}(k)W^{-1}(k,k')\Xi_{\tau s}(k'),
\end{align}
where we have introduced $\Xi_{\tau s}$, $\Delta_{ss'}$, and $\bar{\Delta}_{ss'}$ for decoupling the intra-valley $u$-channel and the inter-valley BCS channel, respectively. Altogether with $\mathcal{S}_0$ term, we can express our theory in terms of a BdG form. We focus on the pairing channel $\Delta_{ss'}$ in the following.  
\begin{align}
	\nonumber&\mathcal{S}_0+\mathcal{S}_I\\
	\nonumber\rightarrow&\sum_{k}\left[\begin{array}{cc}
		\bar{c}_{+s,k} & c_{-s',-k} 
	\end{array}\right]\!\!
	\left[\!\begin{array}{cc}
		G^{-1}_{+s}(k) & \Delta_{s,s'}(k)\\[2mm] 
		\bar{\Delta}_{s,s'}(k) & G^{-1}_{-s'}(k)
	\end{array}\!\right]\!\!\left[\!\begin{array}{c}
		c_{+s,k}\\[2mm]
		\bar{c}_{-s',-k}
	\end{array}\!\right]\\
	\nonumber&+ \mathcal{\beta}\mathcal{A}\sum_{k,k'}\bar{\Delta}_{s,s'}(k)W^{-1}(k,k')\Delta_{s,s'}(k')\\
	\nonumber&+ \frac{\mathcal{\beta}\mathcal{A}}{2}\sum_{k,k'}\Xi_{+ s}(k)W^{-1}(k,k')\Xi_{+ s}(k')\\
	&+ \frac{\mathcal{\beta}\mathcal{A}}{2}\sum_{k,k'}\Xi_{- s'}(k)W^{-1}(k,k')\Xi_{- s'}(k')
\end{align}
where
\begin{align}
	G^{-1}_{+s}(k)=&-i\omega_n+\mathcal{E}_{+}(\vex{k})-\mu+i\Xi_{+s}(k),\\
	G^{-1}_{-s'}(k)=&-i\omega_n-\mathcal{E}_{-}(-\vex{k})+\mu-i\Xi_{-s'}(-k).
\end{align}

Then, we formally integrate out the Grassmann variable in the imaginary-time path integral and construct a free energy density, given by
\begin{align}
\nonumber\mathcal{F}=&-\frac{1}{\beta\mathcal{A}}\sum_{k}\ln\left[G^{-1}_{+s}(k)G^{-1}_{-s'}(k)-\left|\Delta_{ss'}(k)\right|^2
	\right]\\
\nonumber&+\sum_{k,k'}\bar{\Delta}_{ss'}(k)W^{-1}(k,k')\Delta_{ss'}(k')\\
\nonumber&+\frac{1}{2}\sum_{\tau}\sum_{k,k'}\Xi_{+ s}(k)W^{-1}(k,k')\Xi_{+ s}(k')\\
&+\frac{1}{2}\sum_{\tau}\sum_{k,k'}\Xi_{- s'}(k)W^{-1}(k,k')\Xi_{- s'}(k').
\end{align}
Now, we are in the position to derive the self-consistent equations. We perform functional derivative of $\mathcal{F}$ with respect to $\Xi_{+ s}(k)$, $\Xi_{- s'}(k)$, and $\Delta_{ss'}(k)$ and obtain
saddle point equations as follows:
\begin{align}
\Xi_{+s}(k')=&\frac{1}{\beta \mathcal{A}}\sum_{k}\frac{W(k',k)iG^{-1}_{-s'}(k)}{G^{-1}_{+s}(k)G^{-1}_{-s'}(k)-|\Delta_{ss'}(k)|^2},\\
\Xi_{-s'}(k')=&\frac{1}{\beta \mathcal{A}}\sum_{k}\frac{W(k',-k)(-i)G^{-1}_{+s}(k)}{G^{-1}_{+s}(k)G^{-1}_{-s'}(k)-|\Delta_{ss'}(k)|^2},\\
\Delta_{ss'}(k')=&\frac{1}{\beta\mathcal{A}}\sum_{k}\frac{-W(k',k)\Delta_{ss'}(k)}{G^{-1}_{+s'}(k)G^{-1}_{-s'}(k)-|\Delta_{ss'}(k)|^2}.
\end{align}
Without loss of generality, we can parametrize the self energies as follows:
\begin{align}
	i\Xi_{+s}(k)=&\left(-Z_k+1\right)i\omega_n+\chi_k,\\
	-i\Xi_{-s}(-k)=&\left(-Z_k+1\right)i\omega_n-\chi_k,
\end{align}
where $Z_k$ is the wavefunction renormalization and $\chi_k$ contributes to the dispersion renormalization as well as the quasiparticle life time.

Close to $T_c$, the order parameter $\Delta_{ss'}$ is infinitesimal. Thus, the above equations can be reduced to
\begin{align}
\label{Eq:Eliashberg_1}i\Xi_{+s}(k')=&\frac{1}{\beta \mathcal{A}}\sum_{k}\frac{- W(k',k)}{-iZ_k\omega_n+\mathcal{E}_{+}(\vex{k})-\mathcal{E}_F+\chi_k},\\
\label{Eq:Eliashberg_2}i\Xi_{-s}(k')=&\frac{1}{\beta \mathcal{A}}\sum_{k}\frac{- W(k',k)}{-iZ_k\omega_n+\mathcal{E}_{-}(\vex{k})-\mathcal{E}_F+\chi_k},\\
\label{Eq:Eliashberg_3}\Delta_{ss'}(k')=&\frac{1}{\beta\mathcal{A}}\sum_{k}\frac{W(k',k)\Delta_{ss'}(k)}{\left(Z_k\omega_n\right)^2+\left[\mathcal{E}_{+,b}(\vex{k})-\mathcal{E}_F+\chi_k\right]^2}.
\end{align}

\section{Numerical extracted $T_c$ from Eliashberg theory}\label{App:Elia_numerics}

\begin{figure}[t!]
	\includegraphics[width=0.45\textwidth]{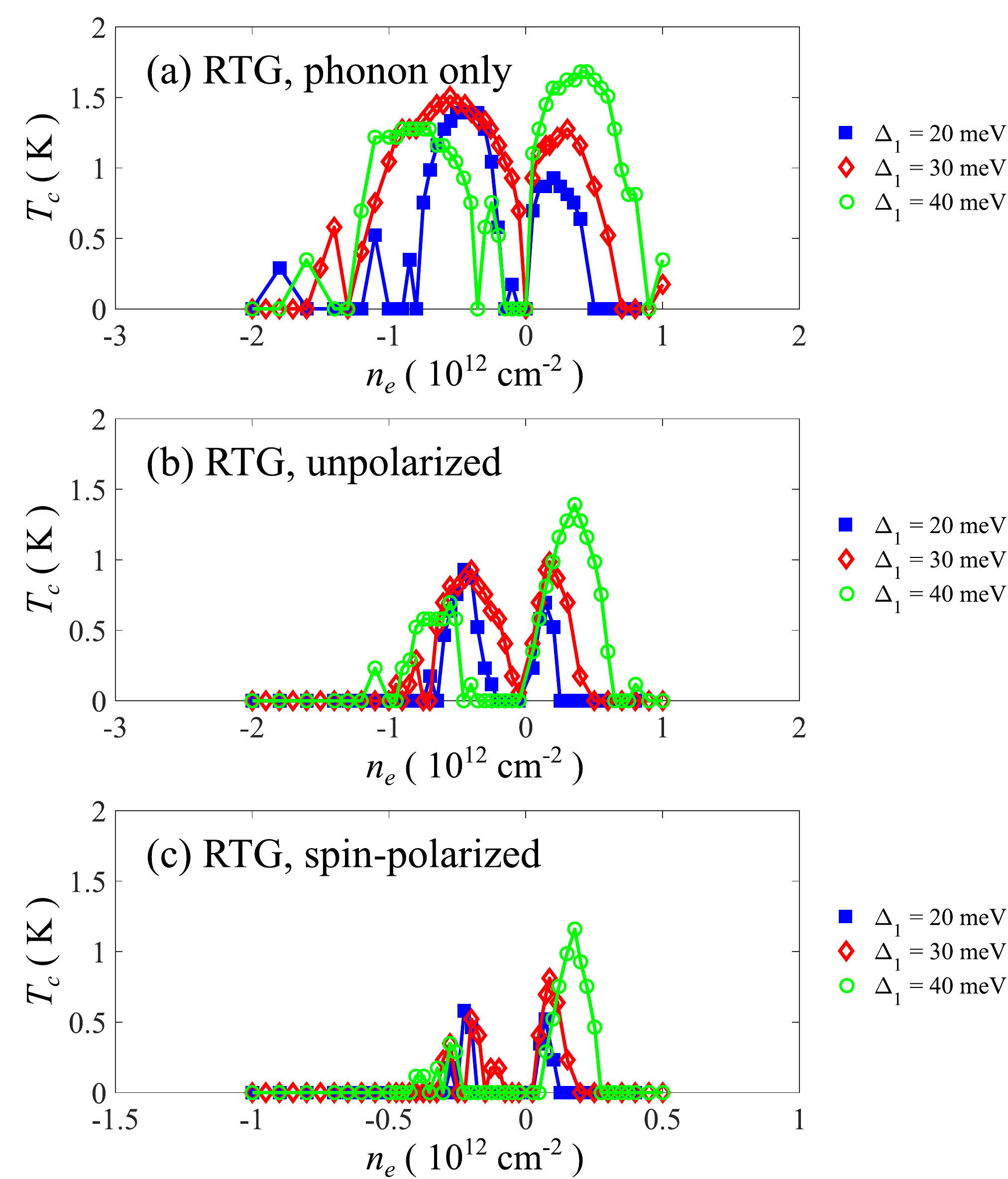}\label{App:Gamma}
	\caption{Superconducting $T_c$ in RTG based on Eliashberg theory [Eq.~(\ref{Eq:LGE_freq})]. (a) Phonon-mediated attraction only. (b) Superconductivity from unpolarized normal states. (c) Superconductivity from spin-polarized normal states. We use a dielectric constant $\epsilon=10$ and a gate distance parameter $d=20$nm for all the data. $\Delta_1$ is a band parameter (defined in Appendix~\ref{App:Bands}) which can be tuned by a displacement field. 
	}
	\label{Fig:Eliash}
\end{figure}

In this appendix, we show numerical results based on Eq.~(\ref{Eq:LGE_freq}), which is the Eliashberg linearized gap equation without including the self-energy corrections. We consider the RTG band and focus on three cases: (a) Phonon-mediated attraction only. (b) Unpolarized normal states. (c) Spin-polarized normal states. The results are summarized in Fig.~\ref{Fig:Eliash}. Despite the quantitative differences between Fig.~\ref{Fig:Eliash} and results in the main text, they all show the same qualitative features, i.e., observable superconductivity prevails for a wide range of dopings. Here, we use a rather small momentum mesh with $\Delta k\approx 0.015 a_0^{-1}$ and keep 100 low-energy levels as well as 100 Matsubara frequencies. We test the numerical procedure with 200 Matsubara frequencies and find essentially the same results. At the lowest temperature, $T=0.1$K, the frequency cutoff $\Lambda$ is around 5meV in our numerical calculations. (Note that the lowest temperature is 10mK for results discussed in main text.) We confirm that the gap function has a Lorentzian-like profile in frequency for case (a). For case (b) and (c), the gap functions, which look like shifted Lorentzian functions, change signs at some particular frequency. The sign changing feature is consistent with the general expectation for Eliashberg theory \cite{Coleman2015introduction,Rietschel1983}.

\section{Evaluating $\Gamma$}

The function $\Gamma(\mathcal{E}_F;\omega_c;\Lambda)$ in Eq.~(\ref{Eq:U_R}) can be evaluated as follows:
\begin{align}
	\Gamma(\mathcal{E}_F;\omega_c;\Lambda)=\frac{1}{\beta\mathcal{A}}\sum_{\substack{\omega_n,\vex{q},\\
			\omega_c<|\omega_n|<\Lambda,\\
			|\tilde{\mathcal{E}}_{\vex{p}}|<\Lambda}}\frac{1}{\omega_n^2+\tilde{\mathcal{E}}_{\vex{p}}^2}.
\end{align}
To simplify $\Gamma(\mathcal{E}_F;\omega_c;\Lambda)$, we consider the zero-temperature limit (i.e., $\beta\rightarrow \infty$) and derive
\begin{align}
	\nonumber&\Gamma(\mathcal{E}_F;\omega_c;\Lambda)=\frac{2}{\mathcal{A}}\sum_{\vex{p},|\tilde{\mathcal{E}_{\vex{p}}}|<\Lambda}\int_{\omega_c}^{\Lambda}\frac{d\omega}{2\pi}\frac{1}{\omega^2+\tilde{\mathcal{E}}_{\vex{p}}^2}\\
	=&\frac{1}{\mathcal{A}}\sum_{\vex{p},|\tilde{\mathcal{E}_{\vex{p}}}|<\Lambda}\frac{1}{\pi|\tilde{\mathcal{E}_{\vex{p}}}|}\left[-\tan^{-1}\left(\frac{|\tilde{\mathcal{E}}_{\vex{p}}|}{\Lambda}\right)+\tan^{-1}\left(\frac{|\tilde{\mathcal{E}}_{\vex{p}}|}{\omega_c}\right)\right].
\end{align}
The above expression can be efficiently evaluated numerically, and $\Gamma(\mathcal{E}_F;\omega_c;\Lambda)$ converges for the mesh sizes used in this work.

\section{Effect of intervalley scatterings on spin-polarized superconductivity}\label{App:Intervalley}
In this section, we discuss how spin-polarized superconductivity is suppressed by intervalley scatterings.
First of all, we consider a spin-polarized spin-triplet superconductor described by
\begin{align}
\mathcal{H}=\sum_{\vex{k}}\Psi^{\dagger}(\vex{k})\hat{h}_{\text{BdG}}(\vex{k})\Psi(\vex{k}),
\end{align}
where $\Psi^{T}(\vex{k})=\left[ c_{+,\vex{k}}; c^{\dagger}_{-,-\vex{k}}\right]$ and
\begin{align}
\hat{h}_{\text{BdG}}=\left[\begin{array}{cc}
	\mathcal{E}_{+}(\vex{k})-\mathcal{E}_F & \Delta(\vex{k})\\[2mm]
	\Delta^*(\vex{k}) & -\mathcal{E}_{-}(-\vex{k})+\mathcal{E}_F
\end{array}
\right],
\end{align}
where $\Delta_{\vex{k}}$ is the pairing potential.
Note that the spin indices are suppressed in the above expression because spins are polarized to the same direction. The system obeys the spinless time-reversal symmetry. As a consequence, $\mathcal{E}_+(\vex{k})=\mathcal{E}_-(-\vex{k})$ is satisfied.

Although the untwisted graphene multilayer systems are clean, the intervalley scatterings (with large momentum transfer) can take place at the sample boundary. In principle, one can study superconductivity with an open boundary condition that exactly facilitates intervalley scattering at the sample termination. However, this is a complicated task as one has to check multiple configurations of the terminated boundary \cite{Akhmerov2008}. To simplify the calculations, we treat the intervalley scattering as a potential in the bulk, and we treat the potential scattering perturbatively upto the second order. The intervalley scattering can be expressed by
\begin{align}
\mathcal{V}=\frac{1}{\sqrt{\mathcal{A}}}\sum_{\vex{k},\vex{k}'}\left[V_{\vex{k},\vex{k}'}c^{\dagger}_{-,\vex{k}}c_{+,\vex{k}'}+\text{H.c.}\right],
\end{align}
where $V_{\vex{k},\vex{k}'}$ encodes the intervalley scatterings.
Note that the spinless time-reversal symmetry gives rise to a condition that $V_{\vex{k},\vex{k}'}=V^*_{-\vex{k},-\vex{k}'}$.
We can extract the mean intervalley scattering $\tau_s$ within the Born approximation,
\begin{align}
\frac{1}{\tau_s}=\left\langle \frac{1}{\mathcal{A}}\sum_{\vex{p}'}|V_{\vex{k},\vex{p'}}|^2\delta(\mathcal{E}_{+}(\vex{k})-\mathcal{E}_F)\right\rangle_{\vex{k}\in \vex{k}_F},
\end{align}
where we have averaged over $\vex{k}$ for $\vex{k}$ on the Fermi surface.

To incorporate the intervalley scattering to superconductivity, we treat $\mathcal{V}$ perturbatively at second order and construct renormalized Gorkov Green functions as follows:
\begin{align}
\hat{\mathcal{G}}^{-1}(\omega,\vex{k})=\hat{\mathcal{G}}^{-1}_0(\omega,\vex{k})-\hat{\Xi}(\omega,\vex{k}),
\end{align}
where $\hat{\mathcal{G}}$ is the renormalized Gorkov Green function, $\hat{\mathcal{G}}_0$ is the bare Gorkov Green function, and $\hat{\Xi}$ is the self energy due to intervalley scattering. The inverse bare Gorkov Green function is expressed by
\begin{align}
	\hat{\mathcal{G}}^{-1}_0(\omega,\vex{k})=\left[\begin{array}{cc}
		\omega -\mathcal{E}_{+}(\vex{k})+\mathcal{E}_F & -\Delta(\vex{k})\\[2mm]
		-\Delta^*(\vex{k}) & \omega+\mathcal{E}_{-}(-\vex{k})-\mathcal{E}_F
	\end{array}
	\right],
\end{align}
and the self energy is expressed by
\begin{align}
	\hat{\Xi}(\omega,\vex{k})=\sum_{\vex{p}}\left[\begin{array}{cc}
		\Xi_{++}(\omega,\vex{k}) & \Xi_{+-}(\omega,\vex{k})\\[2mm]
		\Xi_{-+}(\omega,\vex{k}) & \Xi_{--}(\omega,\vex{k})
	\end{array}
	\right],
\end{align}
where
\begin{align}
	\Xi_{++}(\omega,\vex{k})=&\frac{1}{\mathcal{A}}\sum_{\vex{p}}|V_{\vex{k},\vex{p}}|^2\frac{\omega+\mathcal{E}_{+}(\vex{p})-\mathcal{E}_F}{\omega^2-|\mathcal{E}_{+}(\vex{p})-\mathcal{E}_F|^2-|\Delta(\vex{p})|^2},\\
		\Xi_{+-}(\omega,\vex{k})=&-\frac{1}{\mathcal{A}}\sum_{\vex{p}}\frac{V_{\vex{k},\vex{p}}V_{-\vex{k},-\vex{p}}^*|\Delta(\vex{p})|^2}{\omega^2-|\mathcal{E}_{+}(\vex{p})-\mathcal{E}_F|^2-|\Delta(\vex{p})|^2},\\
			\Xi_{-+}(\omega,\vex{k})=&-\frac{1}{\mathcal{A}}\sum_{\vex{p}}\frac{V_{-\vex{k},-\vex{p}}V_{\vex{k},\vex{p}}^*|\Delta(\vex{p})|^2}{\omega^2-|\mathcal{E}_{+}(\vex{p})-\mathcal{E}_F|^2-|\Delta(\vex{p})|^2},\\
		\Xi_{--}(\omega,\vex{k})=&\frac{1}{\mathcal{A}}\sum_{\vex{p}}|V_{\vex{k},\vex{p}}|^2\frac{\omega-\mathcal{E}_{+}(\vex{p})-\mathcal{E}_F}{\omega^2-|\mathcal{E}_{+}(\vex{p})-\mathcal{E}_F|^2-|\Delta(\vex{p})|^2}.
\end{align}

To simplify the expression of the self energy, we adopt a number of approximations used in Ref.~\cite{Sau2012}. Specifically, we assume that Fermi surfaces are circularly symmetric, $V_{\vex{k},\vex{k}'}$ depends only on the relative angle between $\vex{k}$ and $\vex{k}'$, $\Delta(\vex{k})=\Delta$, and a constant DOS in the integrated energy range. With the above approximations, we can express $\mathcal{G}^{-1}$ as follows:
\begin{align}
\nonumber&\mathcal{G}^{-1}(\omega,\vex{k})\\
	=&a_0(\omega,\vex{k})\hat{1}_{\kappa}+a_z(\omega,\vex{k})\hat\kappa_z+a_+(\omega,\vex{k})\hat\kappa_++a_-(\omega,\vex{k})\hat\kappa_-,
\end{align}
where $\hat{1}_{\kappa}$ is an identity matrix on the Nambu space, and $\kappa_{x,y,z}$ represents the Pauli matrices in the Nambu space, $\kappa_{\pm}=(\kappa_x\pm i\kappa_y)/2$, and
\begin{align}
	a_0(\omega,\vex{k})\approx& \omega\left[1+\frac{\alpha_0}{\tau_s}\frac{1}{\sqrt{|\Delta|^2-\omega^2}}\right],\\
	a_z(\omega,\vex{k})\approx&-\mathcal{E}_+(\vex{k})+\mathcal{E}_F,\\
	a_+(\omega,\vex{k})\approx&\Delta\left[1+\frac{\alpha_1}{\tau_s}\frac{1}{\sqrt{|\Delta|^2-\omega^2}}\right],\\
	a_-(\omega,\vex{k})=&\left[a_+(\omega,\vex{k})\right]^*
\end{align}
In the above expression, $\alpha_0$ and $\alpha_1$ are constants encoding the structure of $V_{\vex{k},\vex{k}'}$. With the approximations stated above, one can easily show that $\alpha_0=1$. If we further impose that $\langle V_{\vex{k},\vex{k}'}\rangle_{\vex{k}'\in\vex{k}_F}=0$ (``statistical valley symmetry''), $\alpha_1=1$ would be obtained.

Now, we are in the position to derive the reduction of pairing potential (corresponding to the quasiparticle excitation energy) in the presence of intervalley scatterings. Following Ref.~\cite{Sau2012}, we obtain the similar equation as follows:
\begin{align}
	\omega\left[1+\frac{\alpha_0}{\tau_s}\frac{1}{\sqrt{|\Delta|^2-\omega^2}}\right]=|\Delta|\left[1+\frac{\alpha_1}{\tau_s}\frac{1}{\sqrt{|\Delta|^2-\omega^2}}\right].
\end{align}
Assuming $x=|\Delta|-\omega\ll |\Delta|$ and using $\alpha_0-\alpha_1=\mathcal{C}$, we obtain
\begin{align}
	x\approx \frac{|\Delta|}{2^{1/3}}\left(\frac{\mathcal{C}}{|\Delta|\tau_s}\right)^{2/3}.
\end{align}
The quasiparticle excitation energy $\omega$ equals to the magnitude of the pairing potential in the presence of intervalley scattering, as described by Eq.~(\ref{Eq:Intervalley}).

%%\bibliography{BIB}

\end{document}